\documentstyle[pra,aps,epsfig,floats]{revtex}
\def\h{\leftrightarrow}
\def\v{\updownarrow}
\begin{document}
\title{Self-homodyne tomography of a twin-beam state}
\author{Giacomo M. D'Ariano,\cite{dar} Michael Vasilyev, and Prem Kumar}
\address{Department of Electrical and Computer Engineering,
Northwestern University, Evanston, IL 60208, USA} 
\date{\today}
\twocolumn
\maketitle
\begin{abstract}
A self-homodyne detection scheme is proposed to perform two-mode
tomography on a twin-beam state at the output of a nondegenerate optical
parametric amplifier. This scheme has been devised to improve
the matching between the local oscillator and the signal modes, which
is the main limitation to the overall quantum efficiency in
conventional homodyning. The feasibility of the measurement is
analyzed on the basis of Monte-Carlo simulations, studying the
effect of non-unit quantum efficiency on detection of the correlation and 
the total photon-number oscillations of the twin-beam state.
\end{abstract}
\pacs{PACS numbers: 42.50.Dv}
\vskip2pc
\narrowtext
\section{Introduction}
One of the most significant advances in modern quantum optics is
the theoretical development \cite{vr} and subsequent experimental realization
\cite{radon-exp} of homodyne tomography. This measurement scheme allows one
to reconstruct the density matrix of the quantum state from a set of field
quadratures measured by a balanced homodyne detector. Reconstruction
methods, initially based on approximate inverse Radon transform of the
quadratures histograms, have been enhanced later through exact
algorithms \cite{my1,my2,my3,direct-sampling} that achieve the
measurement of the matrix element by sampling a corresponding pattern
function of the experimental homodyne outcomes (for a review see \cite{Review}).
These algorithms have been proven to be very stable and fast enough to
allow real-time data sampling. For the photon-number
representation, the calculation of the pattern functions has been
greatly improved by means of factorization formulas
\cite{factor} and asymptotic approximations \cite{WKB} for large photon
numbers of the matrix indices. The direct sampling approach has been
implemented experimentally to measure the photon statistics of a
semiconductor laser \cite{raymer95}, and the density matrix of
a squeezed vacuum\cite{exper-onemode2}. 
The success of optical homodyne tomography has stimulated the
development of state-reconstruction procedures for atomic beams
\cite{atobeams}, the experimental determination of the vibrational state 
of a molecule \cite{diatomolecu}, of an ensemble of helium atoms
\cite{exper-freeatoms}, and of a single ion in a Paul trap
\cite{leibfried}. Finally, some non-tomographic state reconstruction
methods have also been recently proposed \cite{others}.
\par While the full density matrix reconstruction requires the knowledge of the
phase of the detected mode with respect to the local oscillator (LO),
for the diagonal matrix elements it is just sufficient to average over
a random phase \cite{raymer95}. The typical nonclassical states of
interest---such as squeezed states---already exhibit interesting quantum
features in just the photon number distribution; this makes
homodyne tomography especially attractive. Among the quantum
features of interest, there are the even-odd oscillations in the photon 
number distribution of a squeezed vacuum \cite{oscillations1}, which were
recently observed experimentally \cite{exper-onemode2}. In two-mode
tomography of a twin-beam state produced in parametric down-conversion, we are
interested in features of the joint photon-number
distribution of the signal and the idler, such as the delta-function 
correlation
between the photon numbers of the two modes, and the even-odd oscillations
of the total photon number. The sampling algorithm for the two-mode
tomography is obtained by a straightforward extension of the single-mode
case \cite{two-mode}. In the relatively new field of multimode
tomography, recent advances have been made in the theoretical description
\cite{corr-theory} and the experimental measurement \cite{corr-exp} of
the photon-number correlation between two temporal modes. 
\par From the experimental point of view, homodyne tomography of the photon-number
distribution is a viable alternative to direct detection. It
allows one to measure very weak photon fluxes---of the order of a
fraction of a photon per measurement time---using high quantum
efficiency fast {\it p-i-n} photodiodes, as compared to the slow and
less efficient avalanche photodiodes used for direct detection.
This convenience, however, comes with its own price tag. 
One encounters the problem of mode matching between the LO and the
detected modes \cite{matched-lo}, determined by their
spatial/temporal overlap, which gives a detrimental contribution to
the overall quantum efficiency. As shown in Ref. \cite{my2}, the
detection of the quantum features is rapidly degraded by less-than-unity
quantum efficiency of the homodyne detector, and the degree of 
degradation rapidly increases for larger photon numbers.

The problem of mode matching becomes especially severe for quantum states
generated in traveling-wave or pulsed experiments, particularly 
those employing the optical-parametric
amplifiers (OPA's). It has been shown that a LO well matched to
a squeezed vacuum can be generated in the same parametric process
\cite{kumar}. For example, in Ref.~\cite{kumar} a polarizationally 
nondegenerate OPA was
used to produce the squeezed vacuum and the matched LO in two
orthogonal polarizations. However, while this approach is justified
for measurements of the squeezing, it cannot be used for density-matrix 
reconstruction, as a tiny leakage of light from the LO
polarization into the signal polarization can easily spoil the signal 
photon-number distribution. 

In this paper, we address the problem of generating a matched LO for
the reconstruction of the density matrix of the output state of
a polarization-and-frequency nondegenerate OPA. In the spirit of
Ref. \cite{kumar}, we develop a concept of self-homodyning that allows
one to create both the LO and the signal in the same OPA. In the direct
detection of the output signal field, a strong mean field at the
central frequency $\omega_{0}$ can serve as a LO for measuring a mode 
that consists 
of two sidebands at $\omega_{0}\pm\Omega$. As we will show in the following, 
the
relative phase between the LO and the two-sideband mode can be varied,
thus allowing homodyne tomography of the latter. 
In this way one can perform the tomographic reconstruction of full joint
density matrix of the signal and idler twin-beam modes. In this paper,
we consider the measurement of the joint photon-number
distribution of these two modes and 
the photon-number distribution of the signal mode alone. For the latter,
a thermal distribution is expected, as seen in recent self-homodyning
experiments \cite{vasilyev97}. We also show that self-homodyning
can be used to measure the photon statistics of the $+45^{\circ}$- and
the $-45^{\circ}$-polarized linear combinations of the signal and idler modes.

From Monte-Carlo simulations we will estimate the experimental 
conditions that are needed to extract the joint photon-number
probability distribution of the twin beams, the photon correlation
between the modes, and the quantum oscillations of the total photon
number. We will show how these quantities can be experimentally measured
for realistic values of quantum efficiency ($\sim$0.9) of the photodiodes
and for reasonable number of data points ($\sim 10^6$). 

In Section \ref{s:theor} we give a detailed theoretical description of
the self-homodyne measurement, relating the measurement of the field
quadratures to the output photocurrents in Subsection
\ref{ss:detector}, and evaluating the joint probability
distribution of the photocurrents in Subsection \ref{ss:out}, in a form 
suitable for
Monte-Carlo simulations, also taking into account the effect of
non-unit quantum efficiency. In Section \ref{s:qht} we briefly review
the exact reconstruction algorithm for quantum tomography,
for one mode only in Subsection \ref{ss:1mode}, and then with
extension to any number of modes in Subsection \ref{ss:manymodes}. In
Subsection \ref{ss:self} we analyze how the two-mode tomography is
achieved through self-homodyne detection.
In Subsection \ref{ss:bare} we introduce the
concepts of the measurement of the ``dressed'' state, often adopted
in experiments,---as opposed to the
``bare'' state, usually assumed by the theorists. In Section 
\ref{s:MC} we present some selected
Monte-Carlo simulations, also for non-unit quantum
efficiency, for both the bare and the dressed states. We will focus attention
on the joint photon-number probability, on the correlation between the
photon numbers of the two modes, and finally on the total photon-number 
probability, which exhibits oscillations typical of the
twin-beam state. Section \ref{s:discussion} concludes the paper with a
discussion of the results in view of the feasibility of the real
experiment. The Appendix covers the details of derivation of the joint
photocurrent distribution used in Subsection \ref{ss:out}. 
\section{Theoretical description of the self-homodyne measurement}\label{s:theor}
\subsection{The detector\label{ss:detector}}
\begin{figure}[hbt]\begin{center}
\epsfxsize=1\hsize\leavevmode\epsffile{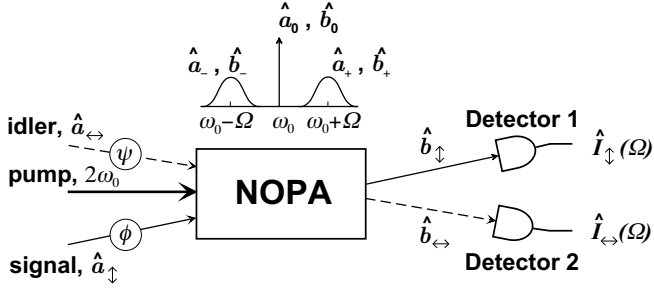}\end{center}
\caption{Scheme of a self-homodyne detector along with the
relevant modes of the electromagnetic field involved in the
measurement. The nondegenerate optical parametric amplifier (NOPA)
is seeded with input fields having strong coherent components  
at frequency $\omega_0$, and is pumped at the second harmonic
$\omega_p=2\omega_0$. At the output of the amplifier the intensities
of the two different polarizations are separately measured by
photodetectors, and a narrow band of the output photocurrents is
selected, centered around frequency $\Omega\ll\omega_0$.} 
\label{f:scheme}\end{figure}

The scheme of a self-homodyne detector is depicted in Fig. \ref{f:scheme}, 
along with the relevant modes of the electromagnetic field involved in
the measurement. A nondegenerate optical parametric amplifier (NOPA)
is injected with an input field having a strong
coherent component at frequency $\omega_0$ with amplitudes
$\alpha_{\v}$ and $\alpha_{\h}$ depending on the polarization, 
$\v$ denoting the vertical and $\h$ the horizontal polarization, respectively.
The amplifier is pumped at the second harmonic $\omega_p=2\omega_0$
with amplitude $\alpha_p\gg\alpha_{\v}, \alpha_{\h}$, such that the 
pump can be considered as classical and undepleted during the
amplification process. At the output of the amplifier two
photodetectors separately measure the intensities of a couple of
orthogonally-polarized components of the field $\hat{\cal E}_{\v}$ and
$\hat{\cal E}_{\h}$. At the output of the photodetectors, a narrow
band of the photocurrent is selected, centered around frequency
$\Omega\ll\omega_0$ (typically $\omega_0$ is optical/infrared,
whereas $\Omega$ is a radio frequency). In the narrowband
approximation and for radiation absorbed in a thin detector layer,
the filtered output photocurrents are given by the (complex) operators
\begin{eqnarray}
\hat I_{\pi}(\Omega)&\propto&\int_{-\infty}^{+\infty}dt\,
e^{i\Omega t}:|\hat{\cal E}_{\pi}(t)|^2:\nonumber \\ 
&=&\int_{-\infty}^{+\infty}d\omega\,\hat{\cal E}^-_{\pi}(\omega+\Omega)
\hat{\cal E}^+_{\pi}(\omega)\;,\quad\pi=\{\v,\h\}\;,
\end{eqnarray}
where $::$ denote the customary normal ordering with the (output)
field annihilation-operator components $\hat{\cal E}^+_{\pi}$ on the right and the
creation operators $\hat{\cal E}^-_{\pi}$ on the left, and the subindex $\pi$ 
runs on the two independent polarizations $\v$ and $\h$.
In terms of the annihilation and creation operators $\hat b$ and
$\hat b^{\dag}$ of the relevant output modes one has
\begin{eqnarray}
\hat I_{\pi}(\Omega)=\hat b_{0\pi}^{\dag}
\hat b_{-\pi}+\hat b_{+\pi}^{\dag} \hat b_{0\pi}\;, 
\label{i_ideal}
\end{eqnarray}
where the subindex $0$ refers to the central mode at frequency
$\omega_0$, and $\pm$ refer to the sidebands at frequencies
$\omega_0\pm\Omega$, respectively. 
The input-output Heisenberg evolutions of the relevant field modes 
across the NOPA are given by
\begin{eqnarray}
\hat b_{0\v}&=&\mu \hat a_{0\v}+\nu \hat a_{0\h}^{\dag}\;,
\qquad \hat b_{0\h}=\mu \hat a_{0\h}+\nu \hat a_{0\v}^{\dag}\;,\nonumber\\
\hat b_{+\v}&=&\mu \hat a_{+\v}+\nu \hat a_{-\h}^{\dag}\;,
\qquad \hat b_{+\h}=\mu \hat a_{+\h}+\nu \hat a_{-\v}^{\dag}\;,\label{io}\\
\hat b_{-\v}&=&\mu \hat a_{-\v}+\nu \hat a_{+\h}^{\dag}\;,
\qquad \hat b_{-\h}=\mu \hat a_{-\h}+\nu \hat a_{+\v}^{\dag}\;,\nonumber
\end{eqnarray}
where $a$ and $a^{\dag}$ denote the annihilation and creation operators
for the input modes, $\mu=\cosh r$, $\nu=e^{i\theta_p}\sinh r$, $r\propto
L\chi^{(2)}|\alpha_p|$ ($L$ is the amplifier length, $\chi^{(2)}$ is the
effective second-order susceptibility). In the following we put 
$\theta_p=0$, namely we set the pump phase  as the reference phase for
all modes. We assume the mode $\hat a_{0\v}$
to be in a highly excited coherent state with amplitude $\alpha_{\v}$.
For the purpose of measurement of the joint photon-number
distribution, the mode
$\hat a _{0\h}$ will also be assumed in a highly excited coherent state
with amplitude $\alpha_{\h}$. In the case where we are interested in
measuring the photon-number distribution of one beam only, the photocurrent
produced by the `$\h$'-polarized beam can be ignored or the mode $\hat
a_{0\h}$ can be assumed to be in the vacuum state.  
In the process of direct detection, the highly-excited central modes 
$\hat a_{0\v,\h}$ beat
with the $\omega_{0}\pm\Omega$ sideband modes, thus playing the role of the LO of
homodyne and heterodyne detectors. This converts the direct detectors
into self-homodyne detectors whose
experimental outcomes are the measured
values of the following rescaled output photocurrents in the limit of
strong LO's: 
\begin{eqnarray}
\hat\i_{\v}(\Omega)&=&\lim_{|\alpha|\to\infty}\frac{\hbox{Tr}_{\rm LO}[\hat
I_{\v}(\Omega)\hat\rho_{\rm
LO}]}{\sqrt2\eta_{\v}|\mu\alpha_{\v}+\nu\alpha^{\ast}_{\h}|}\;,\nonumber\\
\hat\i_{\h}(\Omega)&=&\lim_{|\alpha|\to\infty}\frac{\hbox{Tr}_{\rm LO}[\hat
I_{\h}(\Omega)\hat\rho_{\rm LO}]}{\sqrt2\eta_{\h}|\mu\alpha_{\h}+
\nu\alpha^{\ast}_{\v}|}\;,\label{rescI}
\end{eqnarray}
where $\eta_{\v}$ and $\eta_{\h}$ denote the quantum efficiencies of the
two photodetectors, $\alpha^{\ast}$ denotes the complex conjugate of
$\alpha$, $\hat\rho_{\rm LO}$ represents the density operator of the
LO state, and $\hbox{Tr}_{\rm LO}$ denotes the partial trace
over the LO modes. 
In Eq.~(\ref{rescI}) $\hat I_{\pi}(\Omega)$ is modified from that in 
Eq.~(\ref{i_ideal}) because of the non-unity quantum efficiencies of 
the two photodetectors. It is given by
\begin{equation}
\hat I_{\pi}(\Omega)=\hat b^{\prime\dagger}_{0\pi}\hat b^{\prime}_{-\pi}
+\hat b^{\prime\dagger}_{+\pi}\hat b^{\prime}_{0\pi}\ ,
\label{i_non-unit}
\end{equation}
where $\hat b^{\prime}_{\gamma\pi}=\sqrt{\eta_{\pi}}\;\hat b_{\gamma\pi}+
\sqrt{1-\eta_{\pi}}\; \hat v_{\gamma\pi}\;$. Here $\hat v_{\gamma\pi}$ for
$\gamma\in\{0,+,-\}$ and $\pi\in\{\v,\h\}$ are independent vacuum-state 
operators accounting for the loss at the three
frequency components of each polarization mode.
For the sake of simplicity, we will assume $\eta_{\v}=\eta_{\h}=1$ 
for the rest of this Subsection. We will take into account the effect of 
non-unity quantum efficiency on the photocurrent probability distribution 
in Subsection~\ref{ss:out}. Thus, using Eqs.~(\ref{io}) and quantum 
efficiency equal to unity, one obtains \cite{poms}
\begin{eqnarray}
\hat\i_{\v}(\Omega)&=&\frac{1}{\sqrt2}\left(e^{-i\phi}\hat b_{-\v}+
e^{i\phi}\hat b_{+\v}^{\dagger}\right)\;,\nonumber\\
\hat\i_{\h}(\Omega)&=&\frac{1}{\sqrt2}\left(e^{-i\psi}\hat b_{-\h}+
e^{i\psi}\hat b_{+\h}^{\dagger}\right)\;,\label{phc}
\end{eqnarray}
where $\phi=\mbox{arg}(\alpha_{\v}+\tau\alpha^{\ast}_{\h})$
is the phase of the mode $\hat b_{0\v}$ relative to that of the pump 
with $\tau=\tanh r$, and analogously
$\psi=\mbox{arg}(\alpha_{\h}+\tau\alpha^{\ast}_{\v})$.
Taking the real part of the photocurrents
at given radio-frequency phases $\xi$ and $\chi$ one has  
\begin{eqnarray}
\mbox{Re}\left[\hat\i_{\v}(\Omega)e^{i\xi}\right]&=&\hat
X_{\phi}(\hat B_{\v}^{(\xi)})\;,\nonumber\\ 
\mbox{Re}\left[\hat\i_{\h}(\Omega)e^{i\chi}\right]&=&\hat
X_{\psi}(\hat B_{\h}^{(\chi)})\;,\label{Res} 
\end{eqnarray}
where the operator $\hat X_{\phi}(\hat c)$ denotes
the quadrature at phase $\phi$ of the mode with annihilation operator
$\hat c$, namely, 
\begin{eqnarray}
\hat X_{\phi}(\hat c)={1\over2}\left(e^{-i\phi}\hat c
+e^{i\phi}\hat c^{\dagger}\right)\;,
\end{eqnarray}
and the operator $\hat B_{\pi}^{(\lambda)}$ is the annihilator of the
polarized output mode  
\begin{eqnarray}
\hat B_{\pi}^{(\lambda)}=\frac{1}{\sqrt2}\left(e^{i\lambda}\hat b_{-\pi}+
e^{-i\lambda}\hat b_{+\pi}\right)\;.\label{compo}
\end{eqnarray}
It is easy to check that the output modes (\ref{compo}) have
corresponding input modes given by
\begin{eqnarray}
\hat A_{\pi}^{(\lambda)}=\frac{1}{\sqrt2}\left(e^{i\lambda}\hat a_{-\pi}+
e^{-i\lambda}\hat a_{+\pi}\right)\;,
\end{eqnarray}
and they are related by the Heisenberg evolutions
\begin{eqnarray}
\hat B_{\v}^{(\lambda)}=\mu \hat A_{\v}^{(\lambda)}+\nu
{\hat A}{{}^{(\lambda)}_{\h}}^{\dagger}\;,\nonumber\\ 
\hat B_{\h}^{(\lambda)}=\mu \hat A_{\h}^{(\lambda)}+\nu
{\hat A}{{}^{(\lambda)}_{\v}}^{\dagger}\;.\label{Heis2}
\end{eqnarray}
By scanning the relative phase $\phi$
between $\hat b_{0\v}$ and the pump mode, one can measure any quadrature
$\hat X_{\phi}(\hat B_{\v}^{(\xi)})$ of the output field.
If the input sideband modes $\hat a_{\pm}$ are in a
state with a completely random phase, such as the vacuum, then the only phase 
reference in the output modes $\hat b_{\pm}$ is the pump phase
$\theta_{p}=0$. In that case, the phase $\phi$ can be easily changed by
delaying all the input fields with respect to the pump field, with no need 
to change the phase
of $\hat a_{0\v}$ separately from the other input modes.

From Eq.~(\ref{compo}) one can recognize that 
there are actually four output modes that commute with each other; hence,
their quadratures could be jointly measured by the self-homodyne
detectors. They are $\hat B_{\v}^{(\xi)}$, $\hat B_{\v}^{(\xi+\pi/2)}$,
$\hat B_{\h}^{(\chi)}$, and $\hat B_{\h}^{(\chi+\pi/2)}$, corresponding to the
``cosine'' and ``sine'' components of the two photocurrents in
Eqs. (\ref{phc}) at phases $\xi$ and $\chi$ respectively. 
The modes $\hat B^{(\lambda)}_{\v}$ and $\hat B^{(\lambda)}_{\h}$
are correlated due to the parametric interaction in Eq.~(\ref{Heis2}). This 
interaction, however, does not couple the modes $\hat B^{(\lambda)}_{\v}$ and 
$\hat B^{(\lambda+\pi/2)}_{\h}$.
\subsection{Photocurrent probability distribution\label{ss:out}}
Since we are interested in using self-homodyne detection to measure 
the quadratures of two correlated modes $\hat B^{(\lambda)}_{\v}$ and 
$\hat B^{(\lambda)}_{\h}$, the radio-frequency phase $\lambda$ can 
always be set to zero by shifting the time origin.
Then, the quadratures $\hat X_{\phi}(\hat B_{\v}^{(0)})$
and $\hat X_{\psi}(\hat B_{\h}^{(0)})$ are jointly measured. In the
following we will use the shorthand notation $\hat 
B_{\pi}\equiv \hat B_{\pi}^{(0)}$ and $\hat X^{\pi}_{\phi}\equiv \hat
X_{\phi}(\hat B_{\pi}^{(0)})$, and analogously for the input modes   
$\hat A_{\pi}\equiv \hat A_{\pi}^{(0)}$. 
For perfect detectors, the joint probability
distribution of the ``cosine'' photocurrents with $\xi=\chi$ in 
Eqs.~(\ref{Res}) coincides with the joint probability distribution of the two
quadratures $\hat X^{\v}_{\phi}$ and $\hat X^{\h}_{\phi}$, namely, 
\begin{eqnarray}
p(x,x';\phi,\psi)=\langle x,x';\phi,\psi|\hat R| x,x';\phi,\psi\rangle\;,
\label{idealxx}
\end{eqnarray}
where $| x,x';\phi,\psi\rangle\doteq
|x\rangle_{\phi}\otimes|x'\rangle_{\psi}$ represents the  
simultaneous eigenvector of the two quadratures $\hat X^{\v}_{\phi}$
and $\hat X^{\h}_{\phi}$ with eigenvalues $x$ and $x'$ in the Fock
space ${\cal H}_{\v}\otimes{\cal H}_{\h}$ of the two modes $\hat B_{\v}$
and $\hat B_{\h}$, respectively; and $\hat R$ denotes their joint 
density operator.
For detectors with non-unit quantum efficiencies
$\eta_{\v}$ and $\eta_{\h}$, the joint probability distribution 
$p_{\eta_{\v}\eta_{\h}}(x,x';\phi,\psi)$ of the photocurrents is 
the convolution  \cite{poms} of the ideal probability in
Eq. (\ref{idealxx}) with Gaussians for each mode of variances 
\begin{eqnarray}
\Delta^2_{\eta_{\pi}}=\frac{1-\eta_{\pi}}{4\eta_{\pi}}\;.\label{vareta}
\end{eqnarray}
In this way, the resulting output probability distribution can be
written in the form
\begin{eqnarray}
&&p_{\eta_{\v}\eta_{\h}}(x,x';\phi,\psi)=
{1\over{2\pi\Delta_{\eta_{\v}}\Delta_{\eta_{\h}} }}\times\nonumber\\
&&\mbox{Tr}\left\{\hat R\exp \left[ -{(x-\hat
X^{\v}_{\phi})^2\over{2\Delta^2_{\eta_{\v}}}} 
-{(x'-\hat X^{\h}_{\psi})^2\over{2\Delta^2_{\eta_{\h}}}}
\right]\right\}\;.\label{Gau}
\end{eqnarray}
For simplicity, in the following we will assume equal quantum
efficiencies $\eta_{\v}=\eta_{\h}\equiv\eta$ for both detectors.
Notice that in the limit of unit quantum efficiency, $\eta\to 1$, one
has $\Delta_{\eta}\to 0$, and the ideal probability in (\ref{idealxx})
is recovered. 
\par
\begin{figure}[hbt]\begin{center}
\epsfxsize=.47\hsize\leavevmode\epsffile{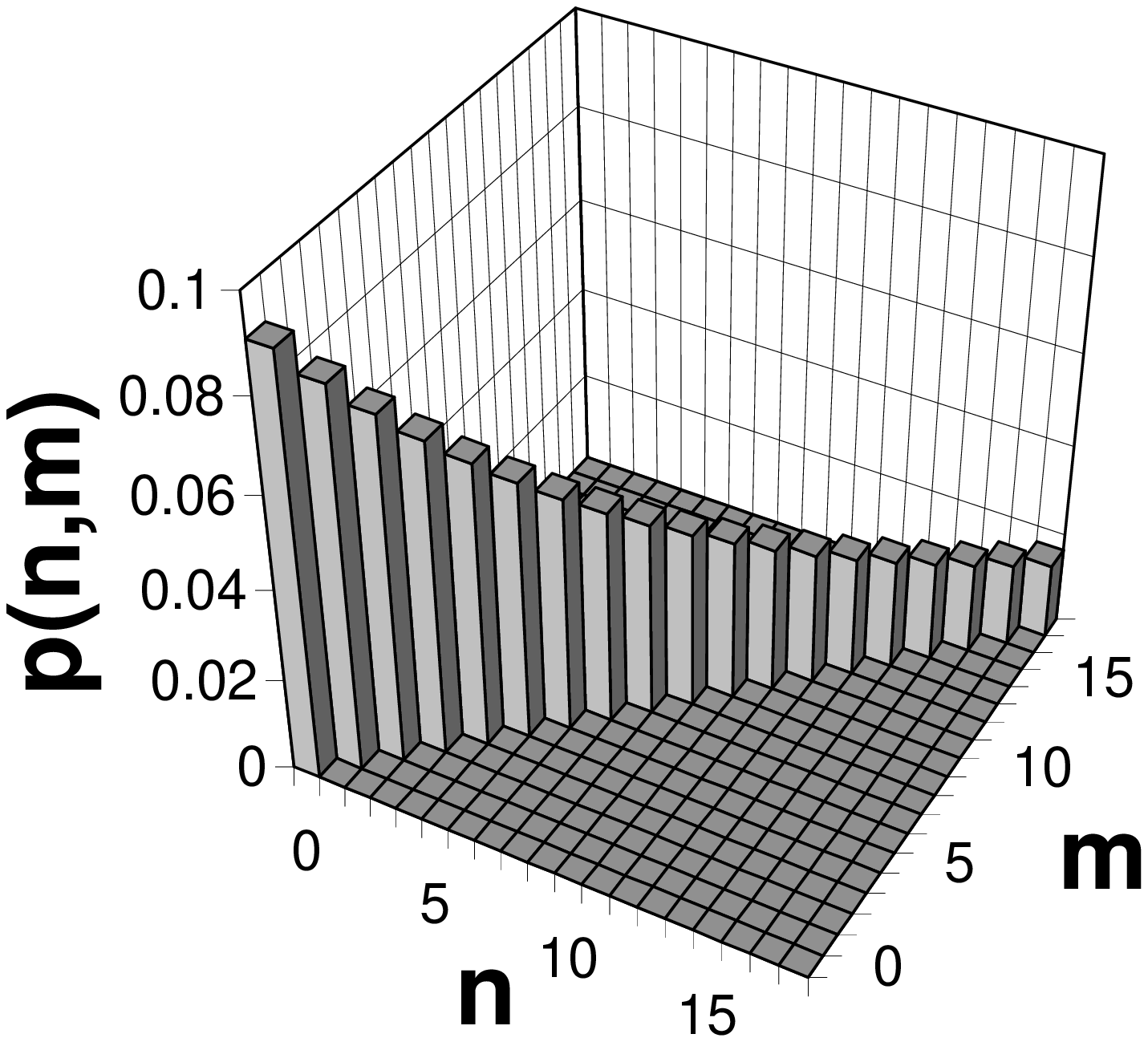}
\epsfxsize=.51\hsize\leavevmode\epsffile{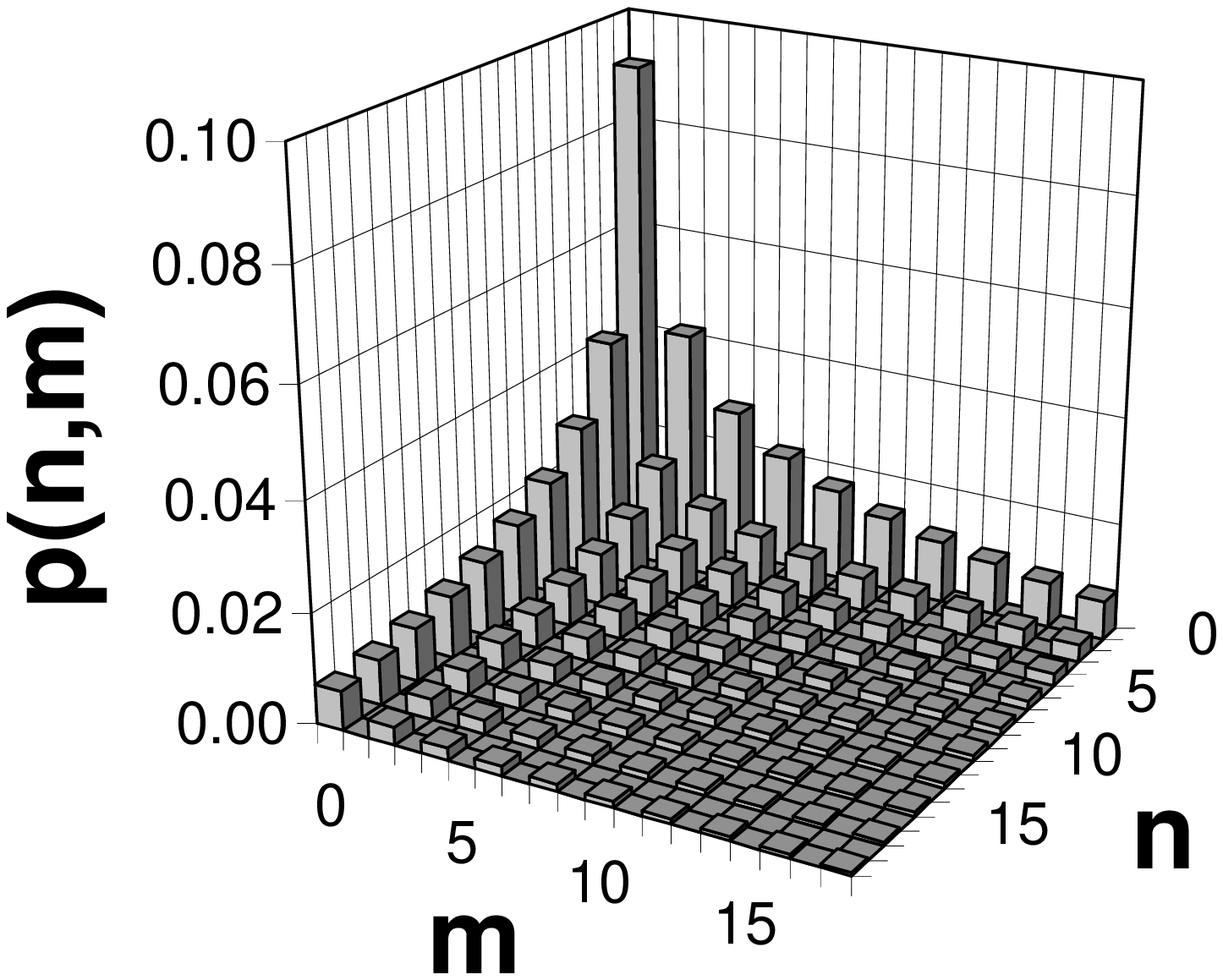}\end{center}
\caption{Theoretical two-mode photon-number probability $p(n,m)$ of
parametric fluorescence in the signal and idler (twin-beam) modes (left),
and in the $\pm 45^{\circ}$-polarized modes (right), given by Eqs.~(\ref{thpn})
and (\ref{p45}), correspondingly. The mean number of
photons in each mode $\bar n=10$.}
\label{f:N1}\end{figure}

 We are now interested in the simplest case of measurement,
that with $\omega_0\pm\Omega$ sidebands in the vacuum state at the input of
the NOPA (i.e., parametric fluorescence). In the Schr\"odinger picture, 
Eqs. (\ref{Heis2})
correspond to the following state generated at the output of the NOPA:
\begin{eqnarray}
|\Psi\rangle=(1-\tau^2)^{1/2}\sum_{n=0}^{\infty}\tau^n|n,n\rangle\;,\label{Psi}
\end{eqnarray}
where $\tau=\tanh r$, and the two-mode Fock
state $|n,m\rangle$ pertaining to $\hat B_{\v}$ and $\hat B_{\h}$ is
given by 
\begin{eqnarray}
|n,m\rangle\doteq
\frac{\left(\hat B_{\v}^{\dagger}\right)^n
\left(\hat B_{\h}^{\dagger}\right)^m}{\sqrt{n!m!}}
|0,0\rangle\;,
\end{eqnarray}
where $|0,0\rangle$ denotes the vacuum for both the $\hat B_{\pi}$
modes.
For the state (\ref{Psi}) the photon-number probability 
is given by 
\begin{eqnarray} 
p(n,m)&\doteq &|\langle n,m|\Psi\rangle|^2\nonumber\\
&=&\delta_{nm}(1-\tau^2)\tau^{2n}={{\delta_{nm}}\over{\bar n +1}}
\left({{\bar n}\over{\bar n +1}}\right)^{n}\;,\label{thpn} 
\end{eqnarray} 
where 
\begin{equation}
\bar n=\frac{\tau^2}{1-\tau^2}=\nu^2=\sinh^2 r\label{barn}
\end{equation}
is the average number of photons in each mode at the NOPA output due
to parametric fluorescence. The main feature of the 
distribution~(\ref{thpn}) is the perfect correlation of the photon numbers in
the signal and idler modes. The two-mode photon-number probability $p(n,m)$ of
Eq.~(\ref{thpn}) is shown in Fig.~\ref{f:N1}(left).
\par
The joint probability distribution of the output photocurrents is
derived in the Appendix, and is given by
\begin{eqnarray}
&&p_{\eta}(x,x';\phi,\psi)\doteq{1\over{2\pi\Delta^2_{\eta}}}\nonumber\\
&\times& \left\langle\Psi\left|\exp
\left\{-{1\over{2\Delta^2_{\eta}}}\left[ (x-\hat X^{\v}_{\phi})^2+
(x'-\hat X^{\h}_{\psi})^2\right]\right\}\right|\Psi\right\rangle\\
&=&{2\over{\pi\sqrt{(d^2_{\kappa}+4\Delta^2_{\eta})(d^2_{-\kappa}+4
\Delta^2_{\eta})}}}\nonumber\\ &\times& \exp\left[
-{{(x+x')^2}\over{d^2_{\kappa}+4\Delta^2_{\eta}}}-
{{(x-x')^2}\over{d^2_{-\kappa}+4\Delta^2_{\eta}}}\right]\;,\label{pxy}
\end{eqnarray} 
which can also be cast in the equivalent form 
\begin{eqnarray}
p(x,x';\phi,\psi)=\frac{a_{\kappa} b_{\kappa}}{\pi}
\exp\left[-a_{\kappa}^2(x-c_{\kappa}x')^2-b_{\kappa}^2{x'}^2
\right]\;,\label{equiv}
\end{eqnarray}
where 
\begin{eqnarray}
\kappa&=&e^{-i(\phi+\psi)}\tanh r\;,\nonumber\\
d^2_{\kappa}&=&{{|1+\kappa|^2}\over{1-|\kappa|^2}}\;,\nonumber\\
a^2_{\kappa}&=&\frac{d^2_{\kappa}+d^2_{-\kappa}+
8\Delta^2_{\eta}}{(d^2_{\kappa}+4\Delta^2_{\eta})
(d^2_{-\kappa}+4\Delta^2_{\eta})}\;,\nonumber\\
c_{\kappa}&=&\frac{d^2_{\kappa}-d^2_{-\kappa}}{
d^2_{-\kappa}+d^2_{\kappa}+8\Delta^2_{\eta}}\;,\nonumber\\
b^2_{\kappa}&=&a^2_{\kappa}(1-c^2_{\kappa})\;.\nonumber
\end{eqnarray}
In the case that we measure only a single output photocurrent, say
$\hat\i_{\v}(\Omega)$---namely, we ignore the measured value of the
other photocurrent $\hat\i_{\h}(\Omega)$---the self-homodyne detector is
equivalent to a conventional homodyne detector, which measures only
the quadrature $\hat X^{\v}_{\phi}$ of mode $\hat B_{\v}$. The output
probability distribution is given by 
\begin{eqnarray}
p_{\eta}(x;\phi)\doteq\frac{1}{\sqrt{2\pi\Delta^2_{\eta}}}\mbox{Tr}
\left\{\hat\varrho\exp\left[-\frac{(x-\hat X^{\v}_{\phi})^2}{
2\Delta^2_{\eta}}\right]\right\}\;,
\label{single-def}
\end{eqnarray}
where the reduced density operator of the mode $\hat B_{\v}$ is
\begin{eqnarray}
\hat\varrho=\mbox{Tr}_{\h}\left[|\Psi\rangle\langle\Psi|\right]=
\frac{1}{\bar n+1}\left(\frac{\bar n}{\bar
n+1}\right)^{\hat B^{\dagger}_{\v}\hat B_{\v}}\;\label{ther}
\end{eqnarray}
with $\hbox{Tr}_{\h}$ denoting the partial trace over the Hilbert
space of the undetected mode $\hat B_{\h}$. The reduced density
operator of the mode $\hat B_{\v}$ in Eq. (\ref{ther}) is that of a thermal 
state with the photon-number probability
\begin{equation} 
p(n)={{1}\over{\bar n +1}}\left({{\bar n}\over{\bar n +1}}\right)^{n}\;,
\label{thpn1} 
\end{equation}
where the average photon number $\bar n$ is given by Eq.~(\ref{barn}). 
The probability
distribution of the output photocurrent, Eq.~(\ref{single-def}), 
is a Gaussian with variance 
$\Delta^2={1\over2}(\bar n+{1\over2})+\Delta^2_{\eta}$, centered at zero. 
This result of self-homodyning of only
the signal mode has been recently demonstrated 
experimentally \cite{vasilyev97}.

Let us note that, while our analysis is aimed at the measurement of the joint
signal-idler photon distribution, a similar self-homodyning approach can also
be implemented to measure the joint distribution of $\pm 45^{\circ}$-polarized
OPA outputs. In that case, a quadrature of the annihilation operator 
\begin{equation}
\hat B_{\pi}=\mu\hat A_{\pi}+\nu\hat A^{\dag}_{\pi}
\label{b45}
\end{equation}
is detected at a phase arg$(\alpha_{\pi}+\tau\alpha^{\ast}_{\pi})$, 
where the subindex 
$\pi$ runs on the two independent $\pm 45^{\circ}$ polarizations, namely
$\nearrow$ and $\nwarrow$, and $\alpha_{\pi}$ is the coherent-state amplitude
of the corresponding central-frequency component of the input. 
Since the interaction (\ref{b45}) does not couple
the $+45^{\circ}$ and $-45^{\circ}$ modes with each other, the polarization
non-degenerate OPA
is equivalent to two degenerate OPA's. Two-mode joint photon-number 
distribution is just a product of the marginal distributions for each mode, 
and in the case of vacuum-state input sidebands is given by 
\cite{oscillations1} 
\begin{eqnarray}
p(n,m)&=&0\qquad\mbox{for $n=2k+1$ or $m=2l+1$}\ ,\nonumber\\
p(n,m)&=&{{\ (2k-1)!!\ (2l-1)!!\ }\over{\ 2^{k+l}\ k!\ l!\ }}\ 
{{1}\over{\bar n +1}}
\left({{\bar n}\over{\bar n +1}}\right)^{k+l}\nonumber\\
&&\qquad\ \mbox{for $n=2k$, $m=2l$}\ ,
\label{p45}
\end{eqnarray}
where the mean photon number $\bar n$ in each mode is given 
by Eq.~(\ref{barn}).
The probability distribution (\ref{p45}) is shown in Fig.~\ref{f:N1}(right),
next to the signal-idler joint photon-number distribution of 
Eq.~(\ref{thpn}). While the
$\pm 45^{\circ}$ modes exhibit independent photon-number oscillations in 
Fig.~\ref{f:N1}(right), the 
signal and idler correlations in Fig.~\ref{f:N1}(left) result in
oscillations of the total photon number. 
\section{Quantum homodyne tomography}\label{s:qht}
\par In this section we briefly review the method for reconstructing the
quantum state that was introduced in Refs.~\cite{my1,my2} for one
field mode. Then we show how it can be straightforwardly
extended to any number of modes---in particular, to the case of two
modes involved in the self-homodyne detection of the OPA output---and 
we will obtain an
algorithm similar to those in Refs.\cite{two-mode}.
Finally, we introduce the
measurement of the ``dressed'' state, often performed in experiments, as 
opposed to the ``bare'' state, typically assumed by the theorists.
\subsection{Single mode detection\label{ss:1mode}}
\par The method for reconstructing the matrix elements of the density
operator is based on the following resolution of the identity on the
Hilbert-Schmidt space
\begin{equation}
\hat\varrho =\int\frac{d^2w}{\pi}\,\hbox{Tr}
[\hat\varrho \hat D(w)] \hat D^{\dagger}(w)\;,\label{HilSc}
\end{equation}
where the integral is extended to the complex plane ${\bf C}$ for
$w$, and $\hat D(w)=\exp(-w^{\ast} \hat a+w
\hat a^{\dagger})$ denotes the displacement operator for the field mode of
interest with annihilation operator $\hat a$. Equation (\ref{HilSc}) simply
follows from the orthogonality relation for displacement operators 
\begin{eqnarray}
\hbox{Tr}[\hat D(w)\hat D^{\dagger}(v)]=\delta_{2}(w-v)\;,\label{ortho}
\end{eqnarray}
where $\delta_{2}(w)$ denotes the Dirac delta-function on the
complex plane. By changing to polar variables $w = (i/2)k
e^{i\phi}$, Eq. (\ref{HilSc}) becomes
\begin{equation}
\hat{\varrho} = \int^{\pi}_0 \, \frac{d\phi}{\pi}\, \int^{+\infty}_{-\infty}\,
\frac{d k\, |k|}{4}\,\hbox{Tr} (\hat{\varrho}
e^{ik\hat{X}_{\phi}})\,e^{-ik\hat{X}_{\phi}}\;,\label{op}  
\end{equation}
where $\hat X_{\phi}=\frac{1}{2} \left(\hat a^{\dagger} e^{i\phi}+\hat a
e^{-i\phi}\right)$ denotes the quadrature operator 
for the field mode $\hat a$. 
Then we evaluate the trace using the eigenvectors
$\{|x\rangle_{\phi}\}$ of $\hat{X}_{\phi}$, and multiply and divide
the function inside the integral by $\exp[(1-\eta)k^2/(8\eta)]$ in the
following fashion:
\begin{eqnarray}
\hat\varrho&=&\int^{\pi}_0 \, \frac{d\phi}{\pi}\, \int^{+\infty}_{-\infty}\,
\frac{d k\,|k|}{4} e^{-\frac{1-\eta}{8\eta}k^2}\nonumber\\
&\times&\int_{-\infty}^{+\infty}\, 
dx\, p(x,\phi) e^{ikx} e^{\frac{1-\eta}{8\eta}k^2}
e^{-ik\hat{X}_{\phi}}\;,\label{op1}  
\end{eqnarray}
where $p(x,\phi)={}_{\phi}\langle x|\hat\varrho|x\rangle_{\phi}$ is
the ideal homodyne probability. Using the convolution theorem we obtain
\begin{equation}
\hat{\varrho} = \int^{\pi}_0 \frac{d\phi}{\pi} \,
\int^{+\infty}_{-\infty} \, d x\, p_{\eta} (x;\phi)
\hat K_{\eta} (x-\hat{X}_{\phi}) \, ,
\label{p2}
\end{equation}
where $p_{\eta} (x;\phi)$ is the homodyne probability distribution for
non-unit quantum efficiency $\eta$, which is the convolution of
$p(x,\phi)$ with a Gaussian of variance $\Delta^2_{\eta}$ given in
Eq. (\ref{vareta}). The kernel $\hat K_{\eta} (x-\hat X_{\phi})$ in 
Eq. (\ref{p2}) is formally given by 
\begin{eqnarray}
&&\hat K_{\eta} (x-\hat X_{\phi}) =\nonumber\\
&&\frac{1}{2}\ \hbox{Re} \int^{+\infty}_0 \, d k\,
k\,\exp \left[ \frac{1-\eta}{8\eta}k^2+ik(x-\hat X_{\phi})\right] 
\;,\label{kf2}
\end{eqnarray}
where convergence of the integral in Eq. (\ref{kf2}) for the operator
$\hat K_{\eta} (x-\hat{X}_{\phi})$ is intended in the weak sense of 
convergence of the matrix elements 
$\langle\upsilon| \hat K_{\eta} (x-\hat X_{\phi})|\upsilon'\rangle$ 
between the Hilbert-space vectors $|\upsilon\rangle$ and
$|\upsilon'\rangle$, which are evaluated before integration.
From Eq. (\ref{p2}) it follows that the matrix element 
$\langle\upsilon|\hat{\varrho}|\upsilon'\rangle$ can be experimentally
obtained by averaging the function  
$\langle\upsilon|\hat K_{\eta} (x-\hat X_{\phi})|\upsilon'\rangle$ over the
quadrature outcomes $x$ that are homodyne detected at random phases
$\phi$ with respect to the LO, namely,
\begin{eqnarray}
\langle\upsilon|\hat{\varrho}|\upsilon'\rangle=
\overline{\langle\upsilon|\hat K_{\eta} (x-\hat X_{\phi})|\upsilon'\rangle}\;,
\label{exper}
\end{eqnarray}
where the overbar denotes the experimental average. The functions 
$\langle\upsilon|\hat K_{\eta} (x-\hat X_{\phi})|\upsilon'\rangle$ for different
vectors $|\upsilon\rangle$ and $|\upsilon'\rangle$ are called
``pattern-functions'' after Ref. \cite{my3}. 
\begin{figure}[hbt]\begin{center}
\epsfxsize=.49\hsize\leavevmode\epsffile{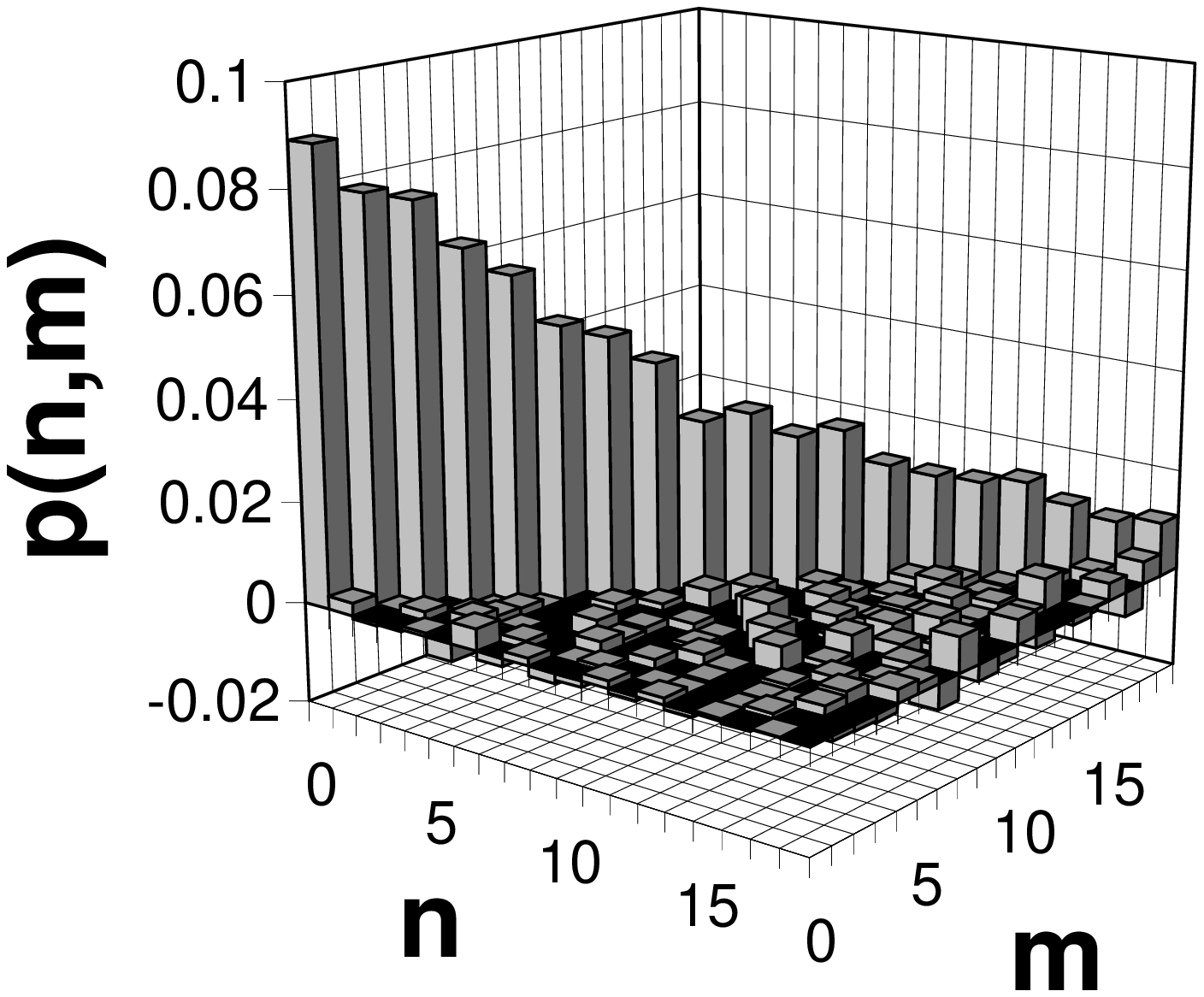}
\epsfxsize=.49\hsize\leavevmode\epsffile{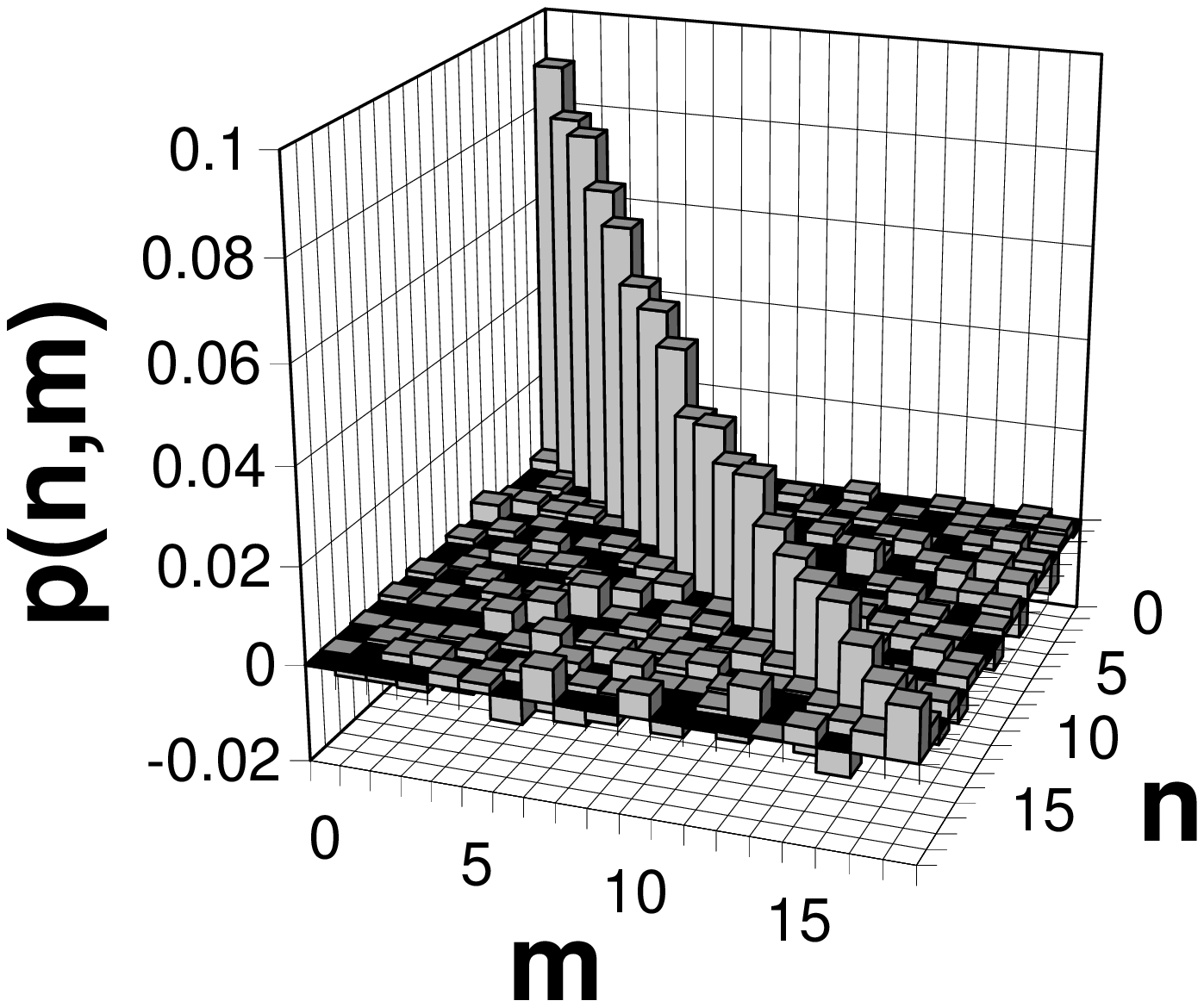}\end{center}
\caption{Two-mode photon-number probability $p(n,m)$ of the twin-beam state
of parametric fluorescence in Eq.~(\ref{Psi}) (two different perspectives),
obtained by a Monte-Carlo simulation of self-homodyne tomography at unit 
quantum
efficiency for $\bar n=10$ and with $10^6$ simulated data.} 
\label{f:A2}\end{figure}
\begin{figure}[hbt]\begin{center}
\epsfxsize=.45\hsize\leavevmode\epsffile{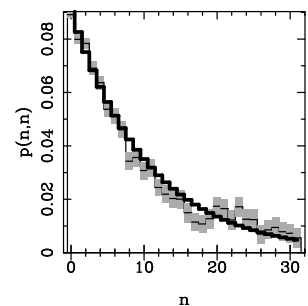}\end{center}
\caption{Diagonal elements $p(n,n)$ of Fig. \ref{f:A2} (shown by thin 
solid line on an extended abscissa range) with their respective error 
bars in gray shade,
compared to the theoretical probability (\ref{thpn}) (thick solid
line).} \label{f:A3}\end{figure}
\par In Ref. \cite{my2}
the boundness of different types of matrix elements of the operator
kernel $\hat K_{\eta} (x-\hat X_{\phi})$ was analyzed as a function of
the  quantum
efficiency. It was shown that for the photon-number and
coherent-state representations these matrix elements become unbounded
for $\eta\le 1/2$. The fact that $\eta=1/2$ is a lower bound for measuring the
state in any (not exotic) representation was thoroughly discussed
in Ref. \cite{Review}. The way in which nonunit quantum efficiency
manifests its detrimental effect when approaching the lower bound is
through increasingly large statistical errors. Let us restrict our
attention to the photon-number representation. At $\eta=1$, as proven
in Ref. \cite{errors}, the statistical errors of the diagonal matrix
elements $\langle n|\hat\varrho|n\rangle$ saturate at the limiting
value $\sqrt{2/N}$ for sufficiently large $n$, independently of the
state $\hat\varrho$ ($N$ is the number of data collected in the
experiment). Also, errors of the off-diagonal elements increase 
very slowly
versus the distance from 
the main diagonal. On the other hand, for $\eta<1$ the errors increase
dramatically versus either $n$ or $1-\eta$, and eventually become
infinite at the lower bound $\eta=1/2$ \cite{comment}. In the next
section we will see how this behavior manifests itself in the two-mode
tomography measurement, on the basis of numerical results from
Monte-Carlo simulation experiments.
\subsection{Multimode detection\label{ss:manymodes}}
\par It is easy to see that Eq. (\ref{HilSc}) can be extended because of
linearity to the case of $M$ modes as follows:
\begin{eqnarray}
\hat R=\int_{{\bf
C}^M}\prod_{l=1}^M\frac{d^2w_l}{\pi}\mbox{Tr}\left[\hat R
\prod_{s=1}^M \hat D_s(w_s)\right]\prod_{r=1}^M \hat D^{\dagger}_r(w_r)\;,
\end{eqnarray}
where $\hat R$ now denotes the joint $M$-mode density operator,
and $\hat D_l(w_l)$ is the displacement operator for the $l$-th
mode. As a consequence, Eq. (\ref{exper}) is extended to the multimode
measurement in the following way:
\begin{eqnarray}
\langle\Phi|\hat R|\Phi'\rangle=
\overline{\langle\Phi|\prod_{l=1}^M \hat K_{\eta_l} 
(x_l-\hat X^{(l)}_{\phi_l})|\Phi'\rangle}\;,
\label{expermulti}
\end{eqnarray}
where $|\Phi\rangle$ and  $|\Phi'\rangle$ are now multimode vectors,
and the experimental average is taken over the random outcomes $x_l$ of
the joint homodyne measurement of quadratures $\hat
X^{(l)}_{\phi_l}$, $l=1,...,M$, of all $M$ modes with random LO phases
$\phi_l$ (we have also let the quantum efficiency to be different for
each homodyne detector). Equation~(\ref{expermulti}) agrees
with the results obtained in Refs.~\cite{two-mode}.
\subsection{Two-mode tomography through self-homodyning\label{ss:self}}
\par As we have seen in Section \ref{s:theor}, in the 
self-homodyne measurement one can jointly measure the quadratures 
$\hat X^{\v}_{\phi}$ and $\hat X^{\h}_{\psi}$ of two different modes 
$\hat B_{\v}$ and $\hat B_{\h}$, and thus, in principle, perform a two-mode
tomography of the OPA output. However, in
order to perform two-mode tomography we need uncorrelated phases
$\phi$ and $\psi$ for the quadratures, whereas in the self-homodyne
measurement they are actually correlated. In fact, one has
\begin{eqnarray}
\phi&=&\mbox{arg}(\alpha_{\v}+\tau\alpha^{\ast}_{\h})\;,\nonumber\\
\psi&=&\mbox{arg}(\alpha_{\h}+\tau\alpha^{\ast}_{\v})\;.
\end{eqnarray}
In the case when we are interested in the photon distribution of one mode
only, we can assume $\alpha_{\h}=0$. Then, by letting the phase of 
$\alpha_{\v}$ fluctuate with a uniform distribution from $0$ to $2\pi$, 
one can perform
one-mode tomography of Eq.~(\ref{exper}). On the other hand, if we are 
interested in the joint photon-number distribution, then we can not make 
the measurement 
by simply averaging
the two-mode pattern functions over the experimental outcomes as in 
Eq.~(\ref{expermulti}). This is because the LO phases
$\phi$ and $\psi$ in this case are not independent random variables. We will
show, however, that it is possible to take the correlation of $\phi$ 
and $\psi$ 
into account and still perform two-mode tomography by appropriately weighting 
the experimental outcomes in Eq.~(\ref{expermulti}).
We first rewrite Eq. (\ref{expermulti}) in the
two-mode case as follows:
\begin{eqnarray}
&&\langle\Phi|\hat R|\Phi'\rangle=
\int_0^{2\pi}\frac{d\phi}{2\pi}\int_0^{2\pi}\frac{d\psi}{2\pi}
\int_{-\infty}^{+\infty}\, dx\,\int_{-\infty}^{+\infty}\, dx'
\nonumber\\
&&\times p(x,x';\phi,\psi)\,\langle\Phi|\hat K_{\eta} (x-\hat
X^{\v}_{\phi})\hat K_{\eta} (x'-\hat X^{\h}_{\psi})|\Phi'\rangle\;.
\label{phaseave} 
\end{eqnarray}
We focus our attention on the phase average only. 
For LO's with equal intensities $|\alpha_{\h}|=|\alpha_{\v}|$ 
and phases $\varphi_{\v}=\mbox{arg}(\alpha_{\v})$ and
$\varphi_{\h}=\mbox{arg}(\alpha_{\h})$, one has 
\begin{eqnarray}
\phi-\psi&=&\varphi_{\v}-\varphi_{\h}\;,\nonumber\\
\phi+\psi&=&\varphi_{\v}+\varphi_{\h}+
2\,\mbox{arg}\left[1+\tau e^{-i(\varphi_{\v}+\varphi_{\h})}\right]\;. 
\end{eqnarray}
After performing the change of variables
\begin{eqnarray}
\sigma&=&{1\over2}(\varphi_{\v}+\varphi_{\h})\;,\nonumber\\
\delta&=&{1\over2}(\varphi_{\v}-\varphi_{\h})\;,
\label{half-sum-diff}
\end{eqnarray}
the average over the phases can be rewritten in terms of the average over
the sum and difference phases with an appropriate weighting function
as follows:
\begin{eqnarray}
\int_0^{2\pi}\frac{d\phi}{2\pi}\int_0^{2\pi}\frac{d\psi}{2\pi}=
\int_{-\pi}^{+\pi}\frac{d\delta}{2\pi}\int_0^{2\pi}d\sigma\,
w(\sigma)\;,
\end{eqnarray}
where the weighting function is given by
\begin{eqnarray}
w(\sigma)=\frac{1}{2\pi}\frac{1-\tau^2}{1+\tau^2+2\tau\cos\sigma}\ .
\label{weight}
\end{eqnarray}
Since the input phases $\varphi_{\v}$ and $\varphi_{\h}$ can certainly
be considered as random and uncorrelated, the same must hold true
for their half-sum $\sigma$ and half-difference $\delta$ in 
Eq.~(\ref{half-sum-diff}). 
Then, the measurement of the matrix element in Eq.~(\ref{phaseave}) is 
obtained by averaging over the experimental random phases $\sigma$ and
$\delta$ with the weighting function~(\ref{weight}). Also,
the weighting function can be rewritten in terms of the gain $g(\sigma)$ of
the central-frequency component that is given by
\begin{equation}
g(\sigma)={{|\mu\alpha_{\v}+\nu\alpha^{\ast}_{\h}|^{2}}\over{|\alpha_{\v}|^{2}}}=
\mu^{2}(1+\tau^{2}+2\tau\cos\sigma )\ .
\end{equation}
Hence, the weighting function is simply
\begin{equation}
w(\sigma)={{1}\over{2\pi g(\sigma)}}\ ,
\label{weight-gain}
\end{equation}
which can be easily and independently measured for every data point
while the homodyne data are collected.

This approach to phase averaging can also be used for detection of the
$\pm 45^{\circ}$ modes mentioned in Subsection~\ref{ss:out}. In that case,
the quadrature phases 
arg$(\alpha_{\nearrow}+\tau\alpha^{\ast}_{\nearrow})$
and arg$(\alpha_{\nwarrow}+\tau\alpha^{\ast}_{\nwarrow})$ are 
independent, but non-uniformly distributed. Then, the averaging is done 
over the input phase $\sigma={\rm arg}(\alpha_{\nearrow})$ or 
$\sigma={\rm arg}(\alpha_{\nwarrow})$, respectively, with the weighting 
function
(\ref{weight-gain}) given by the phase-sensitive gain of the central component.  
\subsection{Measuring the ``bare'' or the ``dressed'' state\label{ss:bare}}
For non-unit quantum efficiency one can measure the density matrix
elements for $\eta$ above the bound $\eta=1/2$. However, instead of
measuring the density matrix of the state $\hat R$ of interest, one
can always measure the density matrix of the state that has been
damped---or ``dressed''---by the quantum efficiency, without any
limitation for $\eta$, even though such a dressed state would be less
and less significant for lower quantum efficiencies. 
The concepts of ``dressed'' and ``bare'' states are two faces of the
same measurement description when regarded in the equivalent
Schr\"odinger and Heisenberg pictures. The conventional description
corresponds to the Heisenberg picture, in which the true state---also
called the ``signal'' or the ``bare'' state---is measured and the effect of
quantum efficiency is ascribed to the detector observable
(photocurrent). In the ``dressed'' state description, on the other hand,
one regards the measurement with $\eta<1$ on the true state $\hat R$
as the corresponding hypothetical ``bare measurement'' with $\eta=1$, but
now on the ``dressed'' state $\hat R_{\eta}$, ascribing the effect of the
non-unit quantum efficiency to the quantum state itself, rather than to
the detector. In other words, the effect of the non-unit quantum efficiency
is regarded in a Schr\"odinger-like picture, with the state evolving
from $\hat R$ to $\hat R_{\eta}$, where the quantum efficiency plays
the role of a time parameter.
\par An easy way to perform tomographic measurement on a
dressed state is just to use the experimental data for $\eta<1$
and analyze them using the pattern function with $\eta=1$. As
shown in Subsection \ref{ss:out}, the effect of non-unit quantum
efficiency is to convolve the quadrature-probability distributions for
all LO phases with a Gaussian of variance $\Delta^2_{\eta}$ given by
Eq.~(\ref{vareta}). This corresponds to convolving the Wigner function
with an isotropic Gaussian of the same variance $\Delta^2_{\eta}$
in the complex plane, which, in turn, corresponds to adding Gaussian noise
to the quantum state. In terms of the bare state $\hat\varrho$, the state
$\hat\Gamma_{\eta}(\hat\varrho)$ dressed with the Gaussian noise is given 
by~\cite{Hall} 
\begin{eqnarray}
\hat\Gamma_{\eta}(\hat\varrho)=\int\frac{d^2w}{\pi\bar{n}}\exp\left(-|w|^2/\bar{m}\right)
\hat D(w)\hat\varrho\hat D^{\dag}(w)\;,\label{Gaussn}
\end{eqnarray}
where the noise-equivalent mean thermal photon number $\bar{m}$ is related to
the quantum efficiency through
\begin{eqnarray}
\bar{m}=2\Delta^2_{\eta}=\frac{1-\eta}{2\eta}\;.\label{mG}
\end{eqnarray}
In the multimode case, one needs to apply the transformation (\ref{Gaussn})
repeatedly, once per each mode, with the corresponding displacement operator 
of the
mode. In the context of measuring the bare state, 
Eq.~(\ref{Gaussn}) was exploited in Ref.~\cite{tokyo} 
to show that the measurement is
possible even in the presence of quantum noise, however, with no more
than $\bar{m}=1/2$ thermal photons.
\begin{figure}[hbt]\begin{center}
\epsfxsize=.48\hsize\leavevmode\epsffile{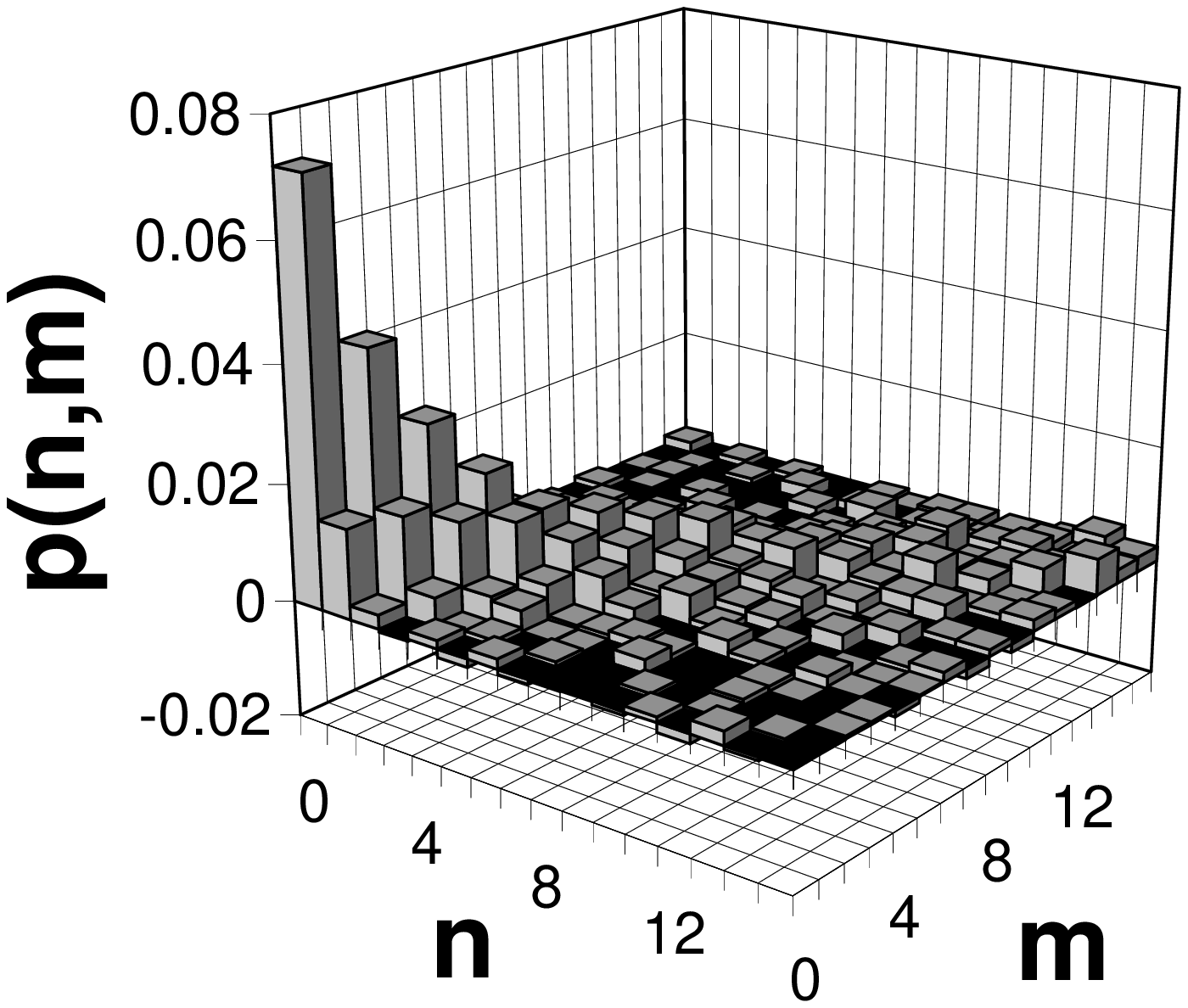}
\epsfxsize=.48\hsize\leavevmode\epsffile{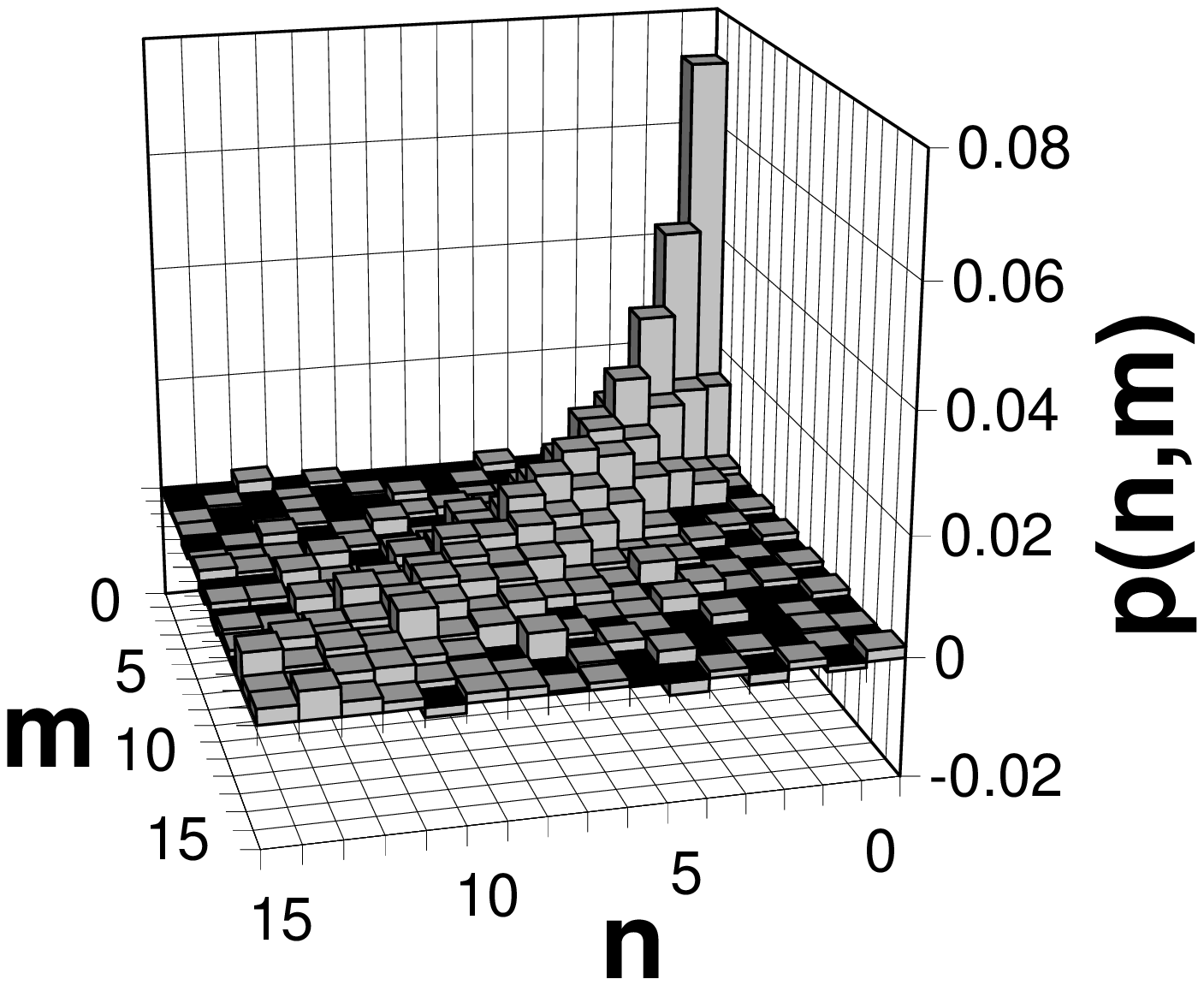}\end{center}
\caption{Monte-Carlo simulation of self-homodyne tomography of the
two-mode photon-number probability $p(n,m)$ for quantum efficiency
$\eta=0.8$ (two different perspectives). The state is the same as 
in Fig.~\ref{f:A2}, but in its 
Gaussian-noise dressed form $\hat\Gamma_{\eta}(\hat R)$ [see
Eq.~(\ref{Gaussn})]. Notice the smearing effect of
the non-unit quantum efficiency.} 
\label{f:B1}\end{figure}
\par Another way of dressing the state, which is often employed
in experimental analysis of the tomographic data (see
Refs.~\cite{raymer95,exper-onemode2,vasilyev97}) is to consider the state
that has undergone a loss equivalent to $\eta$. In this case, the
analysis is done by rescaling the output photocurrents by
$\sqrt{\eta}$ instead of $\eta$ as in Eqs.~(\ref{rescI}), and then
using the pattern functions for $\eta=1$. It is easy to see that this
procedure
corresponds to measuring the dressed state $\hat\Lambda_{\eta}(\hat\varrho)$,
which is related to the bare state $\hat \rho$ as follows:
\begin{eqnarray}
\hat\Lambda_{\eta}(\hat\varrho)=\sum_{n=0}^{\infty}{{(\eta-1)^n}\over{n!}}
\hat a^n \eta^{-{1\over2}\hat a^{\dag}\hat a} \hat\varrho\,
\eta^{-{1\over2}\hat a^{\dag}\hat a} (\hat a^{\dag})^n\;.\label{loss}
\end{eqnarray}
Again, in the multimode case the transformation (\ref{loss}) is
applied separately to all modes. One can also regard the state
$\hat\Lambda_{\eta}(\hat\varrho)$ in Eq.~(\ref{loss}) as the state of the mode
$\sqrt{\eta}\;\hat a+\sqrt{1-\eta}\;\hat v$---instead of the state
of just the mode $\hat a$ of interest---where $\hat v$ is the independent
vacuum-state mode
responsible for the loss. 
\par Before concluding this section, we need to say a few words
regarding the difference between the two dressed states 
$\hat\Lambda_{\eta}(\hat\varrho)$ and $\hat\Gamma_{\eta}(\hat\varrho)$. In the
loss model corresponding to $\hat\Lambda_{\eta}(\hat\varrho)$, the dressed 
state 
loses some signal, and becomes the
vacuum state in the limit of $\eta\to0$, independently of
$\hat\varrho$, which makes the state less and less meaningful for
decreasing $\eta$. On the other hand, in the Gaussian-noise model
corresponding to $\hat\Gamma_{\eta}(\hat\varrho)$, there is
no loss of signal, but the state gets an increasingly large number of
thermal photons for decreasing $\eta$. In this way, the most interesting
quantum features of the state---as, for example, oscillations in the
photon-number probability---are lost, as shown in Ref.~\cite{my2}, and
all states tend to look ``classical''. In the next section we will see
these effects at work in some Monte-Carlo
numerical experiments for the two-mode case.  
\section{Monte-Carlo simulations}\label{s:MC}
In this section we present some numerical results from Monte-Carlo
simulations of the self-homodyne measurement. Our aim is to analyze the
feasibility of a real experiment and to see how many measurements are
needed for a state reconstruction, especially in presence of the
detrimental effect of non-unit quantum efficiency of the
photodetectors. We will restrict our analysis to the measurement of
the joint density matrix $\hat R$ of the two modes, $\hat B_{\v}$
and $\hat B_{\h}$, assumed to be in the correlated state given by 
Eq.~(\ref{Psi}).
\par The simulation of the homodyne outcomes is based on the
probability distribution in Eq.~(\ref{equiv}), which shows how
the outcomes can be obtained from a Gaussian random generator,
starting from the generation of $x'$, then generating $x$, and finally
shifting the latter by $c_{\kappa}x'$. The phases $\phi$ and $\psi$ of the
quadrature are chosen randomly for every sample. The 
density matrix in the photon-number representation is measured by
averaging the pattern functions over the random data:
\begin{eqnarray}
&&\langle n_1,m_1|\hat R|n_2,m_2\rangle=\nonumber\\[2mm]
&&\overline{\langle n_1 |\hat K_{\eta} (x-\hat X_{\phi})|n_2\rangle
\langle m_1 |\hat K_{\eta} (x'-\hat X_{\psi})|m_2\rangle}\;.
\end{eqnarray}
The pattern functions for a generic $\eta$ are obtained from the
pattern functions for $\eta=1$, using the inverse 
generalized Bernoulli transformation as in Ref.~\cite{kiss}.
The pattern functions for $\eta=1$, in turn, are obtained from the
factorization formulas of Refs.~\cite{factor} (following our
conventions for the quadratures, we actually use the factorization
formulas as given in Ref.~\cite{Review}).
In Fig.~\ref{f:A2} we show the results of a simulation for the 
measurement of the 
two-mode photon-number probability $p(n,m)$ for unit quantum
efficiency. The theoretically expected distribution, 
given by Eq.~(\ref{thpn}), is
shown in Fig.~\ref{f:N1}(left). In Fig.~\ref{f:A3} the diagonal
elements $p(n,n)$ of Fig. \ref{f:A2}  are shown with their respective 
error bars, and compared
against the theoretical probability of Eq.~(\ref{thpn}). 
From both Figs.~\ref{f:A2} and \ref{f:A3} we see that there is an
excellent agreement between the theoretically-obtained and
tomographically-reconstructed joint probabilities, and
the fluctuations in the latter are already very small for a 
number of data samples as low as $10^6$, which can be easily acquired 
within the stability time of a typical twin-beam setup.
\begin{figure}[hbt]\begin{center}
\epsfxsize=.45\hsize\leavevmode\epsffile{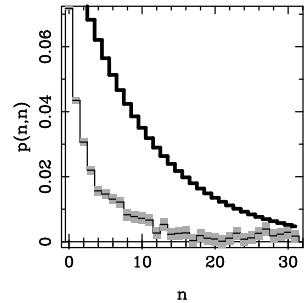}\end{center}
\caption{The same as in Fig.~\ref{f:B1}, but for the diagonal elements 
$p(n,n)$ only (thin solid line with error bars in gray shade), 
compared to the theoretical
probability (\ref{thpn}) for the bare state (thick solid line).
The disagreement between the theoretical probability for the bare state
$\hat R$ and the simulated measurement for the dressed state 
$\hat\Gamma_{\eta}(\hat R)$ is a typical manifestation of the non-unit
quantum efficiency.} 
\label{f:B3}\end{figure} 
\begin{figure}[hbt]\begin{center}
\epsfxsize=.5\hsize\leavevmode\epsffile{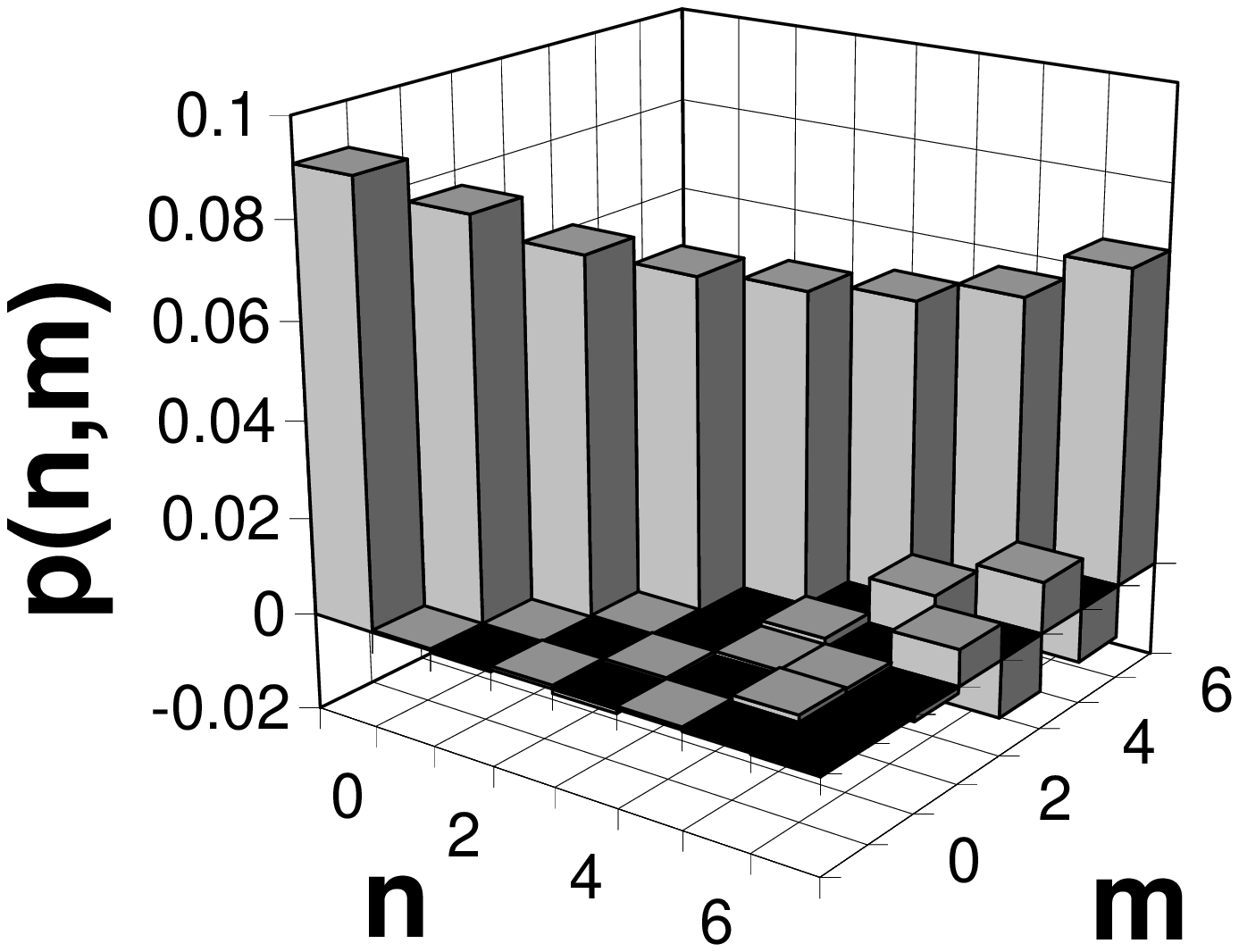}
\epsfxsize=.48\hsize\leavevmode\epsffile{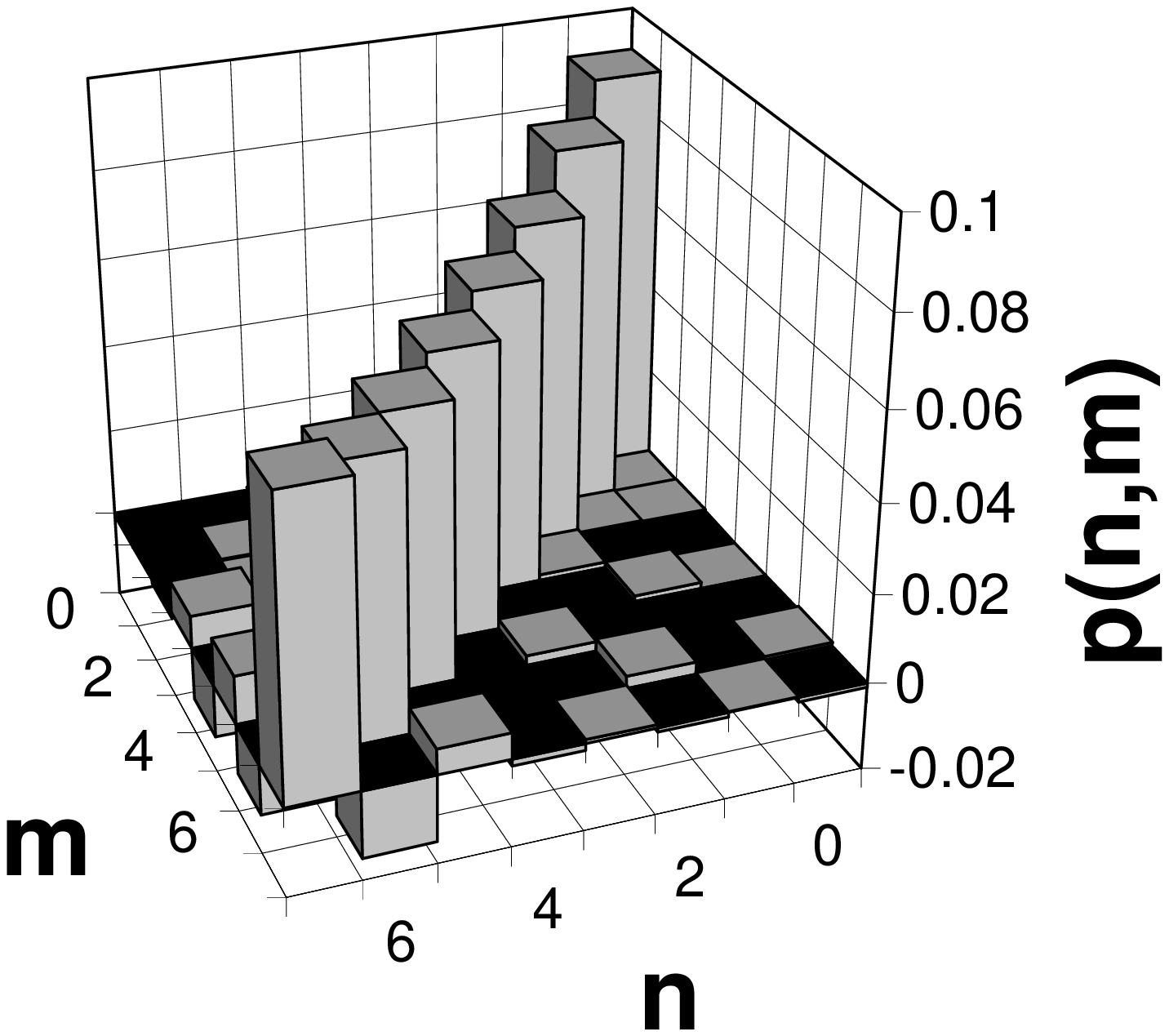}\end{center}
\caption{Reconstruction of the bare state using the pattern
functions with the correct experimental value of the quantum
efficiency $\eta$ (two different perspectives are shown). Here $\bar n=10$,
$\eta=0.9$, and we used $10^7$ data samples 
for the Monte-Carlo simulation.}\label{f:CC1}\end{figure}
\par In Figs.~\ref{f:B1} and \ref{f:B3} the same tomographic measurement of
Figs.~\ref{f:A2} and \ref{f:A3} is reported, but now for  
a quantum efficiency $\eta=0.8$ for each detector. However,
in the reconstruction, the pattern functions for $\eta=1$ are used. 
As explained in Subsection \ref{ss:bare},
this corresponds to a measurement of the state $\hat\Gamma_{\eta}(\hat
R)$ that has been dressed by the Gaussian-noise equivalent of the 
quantum efficiency, 
instead of a measurement of the true twin-beam state. (For values of the
quantum efficiency $\eta=0.8$ and $\eta=0.9$ used throughout this 
paper, the two kinds of state dressing---Gaussian-noise or loss---give
similar qualitative results.) The smearing effect of the
non-unit quantum efficiency is evident in Fig.~\ref{f:B1}, where the
perfect photon-number correlation between the 
two modes is smudged, resulting in non-vanishing probabilities
$p(n,m)$ for $n\neq m$. Because of the preservation of the normalization
in the $(n,m)$ plane, the diagonal $p(n,n)$ is decreased,
resulting in the evident disagreement in Fig.~\ref{f:B3}, where the
reconstructed diagonal elements $p(n,n)$ are reported with relative error 
bars and compared with the theoretical probability (\ref{thpn})
for the bare state.
\begin{figure}[hbt]\begin{center}
\epsfxsize=.45\hsize\leavevmode\epsffile{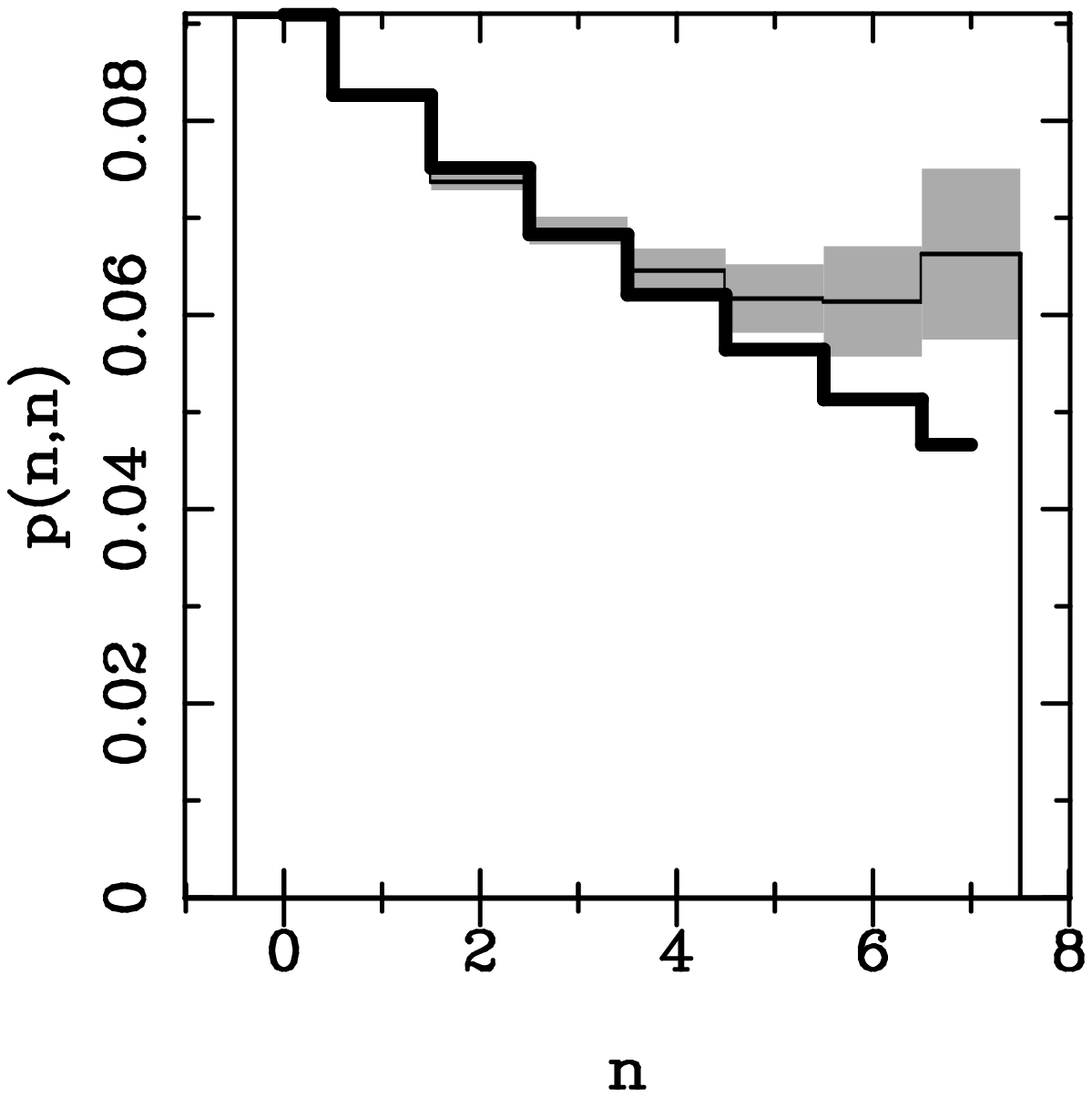}
\epsfxsize=.45\hsize\leavevmode\epsffile{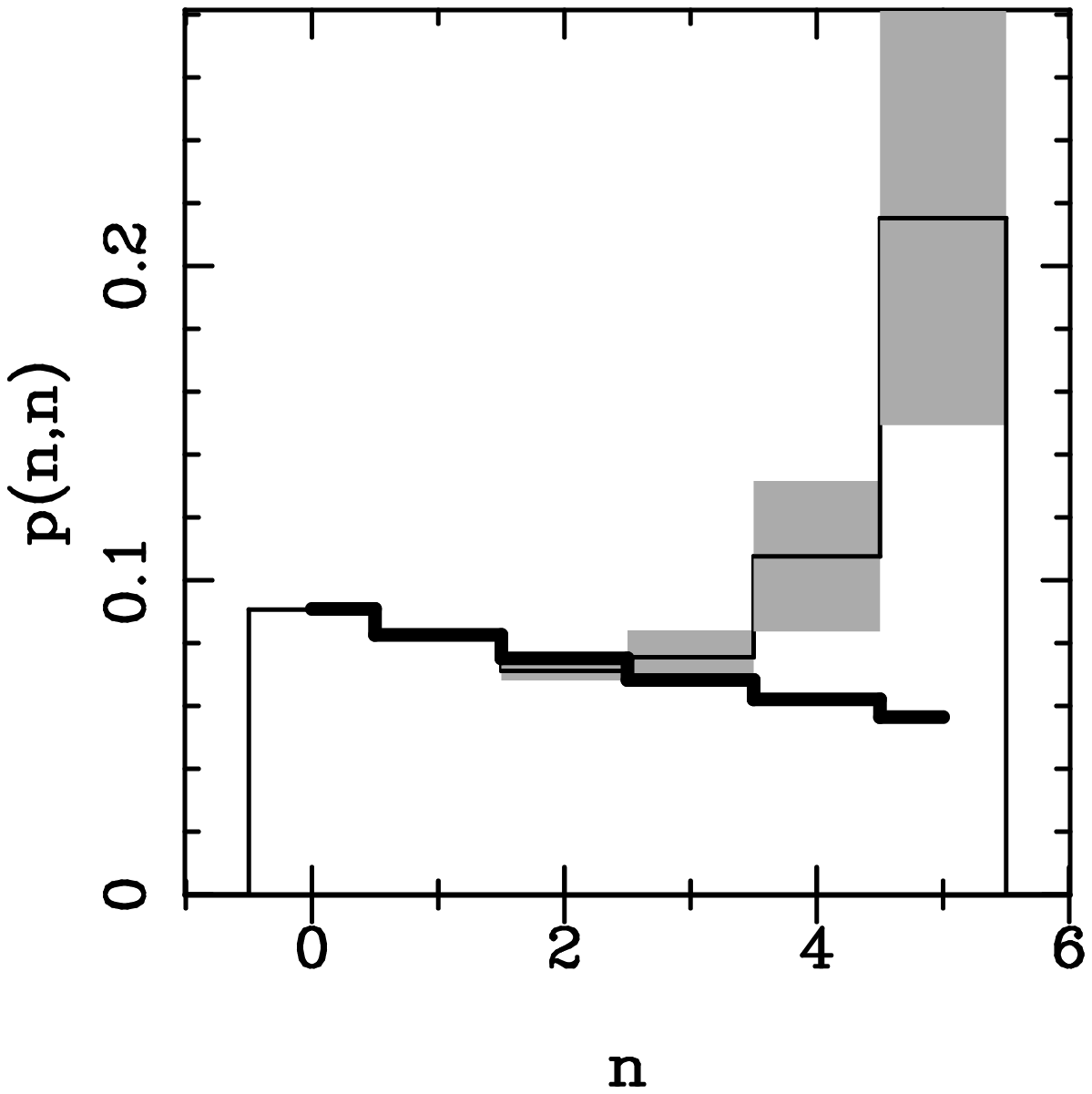}\end{center}
\caption{Reconstruction of the diagonal probability $p(n,n)$ for the
bare state, using the pattern functions with the correct
experimental value of the quantum efficiency $\eta$. Here $\bar n=10$
and $\eta=0.9 (0.8)$ in the left (right) figure.
The theoretical probability (thick solid lines) is
superimposed onto the results of the Monte-Carlo experiments ($10^7$
data samples); the latter are shown with thin solid lines  with statistical 
errors in gray shade. Notice that there is no longer the disagreement
shown in Fig. \ref{f:B3}, but now error bars increase dramatically
versus $n$ and for smaller $\eta$.}\label{f:C3}\end{figure}
\par In Fig.~\ref{f:CC1} we present the results of Monte-Carlo
simulation for a realistic measurement of the bare state, but now using the 
pattern functions with the correct experimental value of the quantum
efficiency $\eta$. One can see that the smearing effect of the non-unit
quantum efficiency has been cleaned out, which, however, comes at the expense 
of increasing
fluctuations for large $n$. This is even more evident in
Fig.~\ref{f:C3}, where the reconstruction of the diagonal probability
$p(n,n)$ for the bare state is shown for two different values, $\eta=0.9$
and $\eta=0.8$, of the quantum efficiency. One can see that there is no
longer the disagreement between the reconstructed and the theoretical 
values, of the kind shown in Fig.~\ref{f:B3}, but now the error bars
have increased dramatically for larger $n$, becoming worse for smaller $\eta$
[cf.~Fig.~\ref{f:C3}(right)].
\begin{figure}[hbt]\begin{center}
\epsfxsize=.45\hsize\leavevmode\epsffile{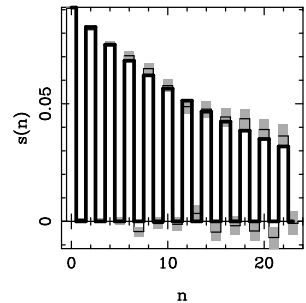}\end{center}
\caption{Oscillations of the total photon-number probability
$s(n)$ in Eq.~(\ref{sum}) due to the perfect correlation of the photon
number in the twin-beam 
state, Eq.~(\ref{Psi}). Thin solid line
with error bars in
gray shade represents the results of a Monte-Carlo simulation 
with unit quantum efficiency, $\bar n=10$, and 
$10^6$ data samples.
Thick solid line is the theoretical result, Eq.~(\ref{ths}).}
\label{f:D1}\end{figure}
\begin{figure}[hbt]\begin{center}
\epsfxsize=.45\hsize\leavevmode\epsffile{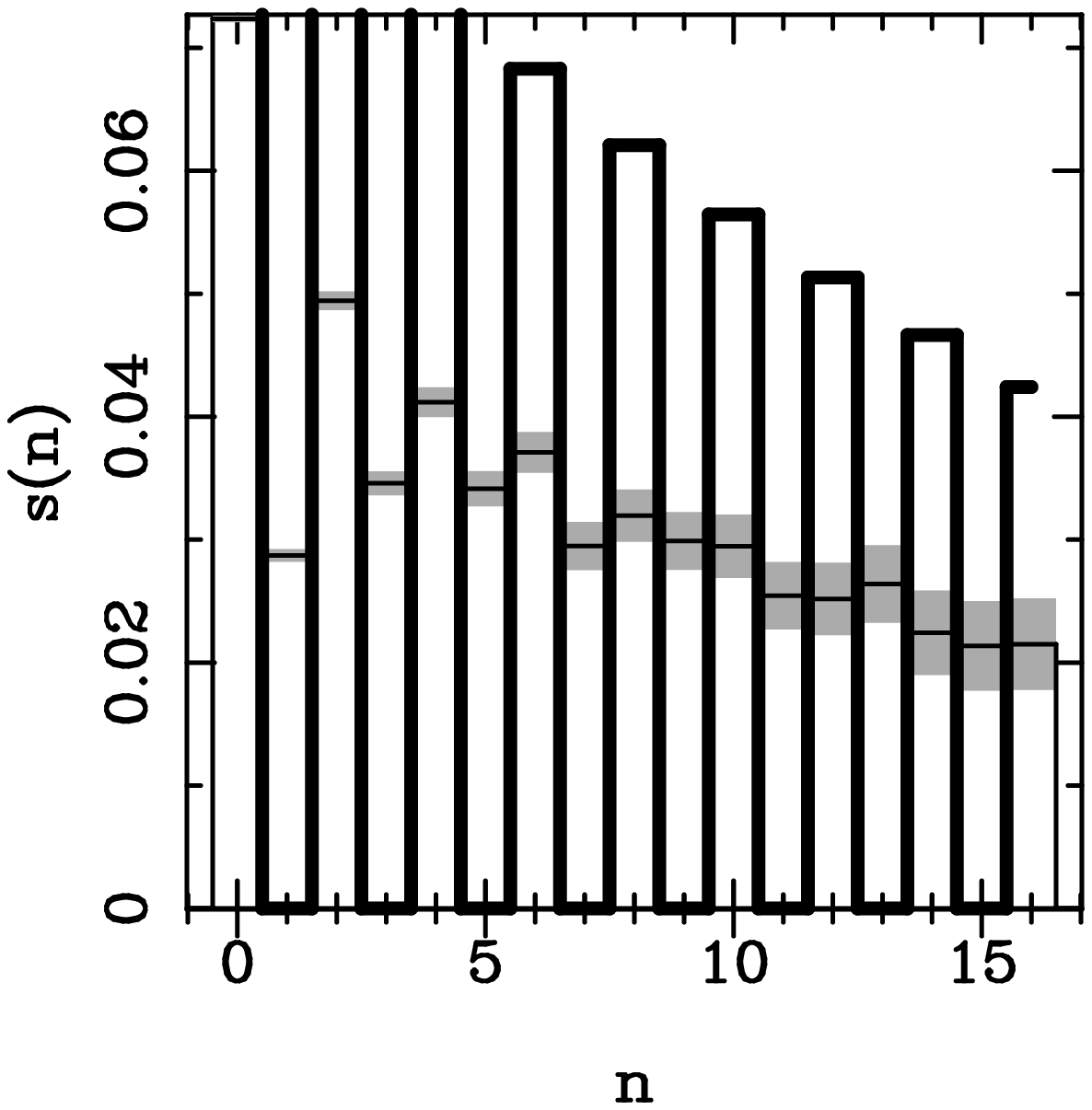}
\epsfxsize=.45\hsize\leavevmode\epsffile{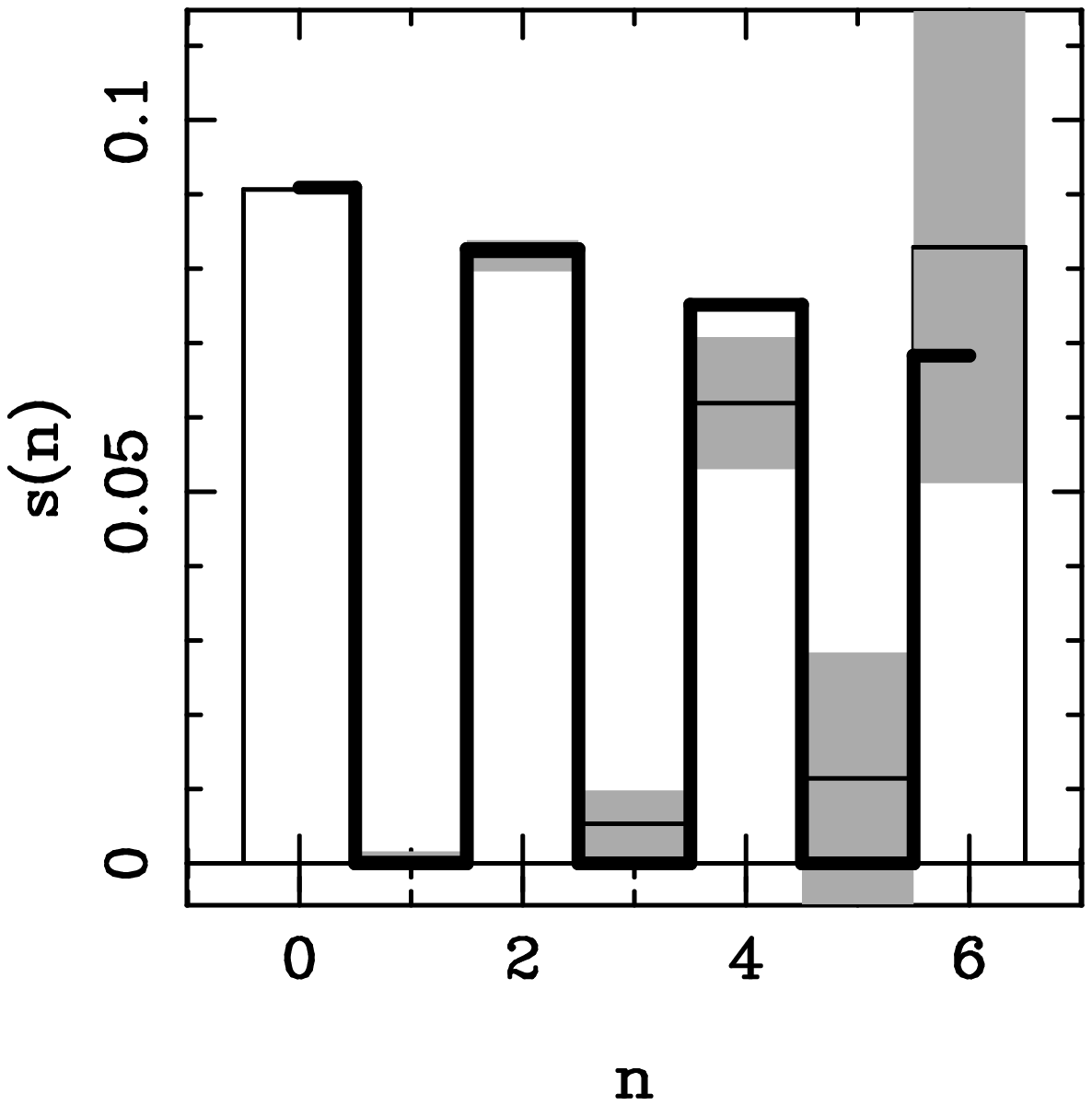}\end{center}
\caption{Similar to Fig. \ref{f:D1}, but for a quantum efficiency $\eta=0.8$.
Results for the Gaussian-noise-dressed state reconstruction 
are shown on the left and for the bare 
state reconstruction on the right. Here $\bar n=10$, and we used 
$5\times 10^6$ data samples for
the left plot and $10^7$ data samples for the right plot. In the left
plot, the ordinate is truncated at the maximum value of the simulated
probability. 
The oscillations are nicely recovered in
the right plot, wherein 
pattern functions with the correct value of quantum efficiency (0.8)
were used for reconstruction..}\label{f:E1}\end{figure}
\begin{figure}[hbt]\begin{center}
\epsfxsize=.45\hsize\leavevmode\epsffile{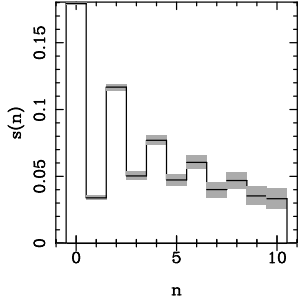}
\end{center}
\caption{Similar to Fig.~\ref{f:E1}(left), but for $\eta=0.9$,
$\bar n=4$, and $10^6$ data samples.}
\label{f:EE1}\end{figure}
\par The off-diagonal number probabilities and the correlation between
the two modes can be analyzed by evaluating the following
sums of matrix elements:
\begin{eqnarray}
s(n)&=&\sum_{l=0}^n \langle l,n-l|\hat R|l,n-l\rangle\;,\label{sum}\\
d_N(n)&=&\sum_{l={\rm max}(-n,0)}^N \langle l,n+l|\hat
R|l,n+l\rangle\;.\label{diff} 
\end{eqnarray}
The quantity $s(n)$ is the probability distribution for the total
number of photons in the two modes. The theoretical result for our state in
(\ref{Psi}) is the oscillating function 
\begin{eqnarray}
s(n)=\cases{(1-\tau^2)\tau^{2n} \quad&n\mbox{ even,}\cr
0\quad&n\mbox{ odd,}\cr}
\label{ths}
\end{eqnarray}
similar to the photon-number distribution of a single-mode squeezed 
vacuum \cite{exper-onemode2,oscillations1}.
On the other hand, the quantity $d_N(n)$ 
represents the photon-number correlation between the two modes, and 
in the limit $N\to\infty$ is the Kroneker $\delta_{n0}$ for a twin-beam state. 
For finite $N$ its theoretical value for the state in Eq.~(\ref{Psi})
can be evaluated to be
\begin{eqnarray}
d_N(n)=\delta_{n0}(1-\tau^{2(N+1)})\;.\label{thd}
\end{eqnarray}
In Fig.~\ref{f:D1} we show the results of a simulation
of the total photon-number probability $s(n)$, Eq.~(\ref{sum}),
for $\eta=1$ and compare them to the theoretical value, Eq.~(\ref{ths}). 
As shown, the theoretically-expected distribution is well reproduced 
from $10^6$ data samples with very small statistical errors. 
In Fig.~\ref{f:E1} a similar simulation
is presented as in Fig.~\ref{f:D1}, but now for a quantum efficiency
of $\eta=0.8$. The total photon-number probability $s(n)$ is 
reconstructed for both the dressed state and the bare 
state. Once again, one can see the smearing effect of the quantum efficiency
in the dressed-state case, where the oscillations of the total photon number
are almost completely washed out. 
On the other hand, the oscillations are nicely recovered in the
reconstruction of the bare state, albeit at the expense of increasingly
large statistical errors. In Fig.~\ref{f:EE1} we present a simulation 
for $\eta=0.9$ to show how these quantum
oscillations would be detected in an experimentally-feasible measurement 
of the dressed state with $\bar n=4$ and $10^6$ data samples.
\par
\begin{figure}[hbt]\begin{center}
\epsfxsize=.45\hsize\leavevmode\epsffile{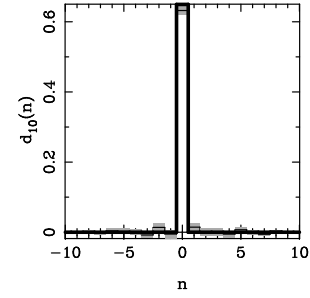}\end{center}
\caption{Correlation function, Eq.~(\ref{diff}), for the twin beam
state in Eq.~(\ref{Psi}) with $\bar n=10$ and $\eta=1$
reconstructed from $3\times 10^5$ data samples. Results of the 
Monte-Carlo simulation
(thin solid line with error bars in
gray shade) are superimposed  onto the theoretical correlation,
Eq.~(\ref{thd}), shown by thick solid line.}\label{f:G1}\end{figure}

\begin{figure}[hbt]\begin{center}
\epsfxsize=.45\hsize\leavevmode\epsffile{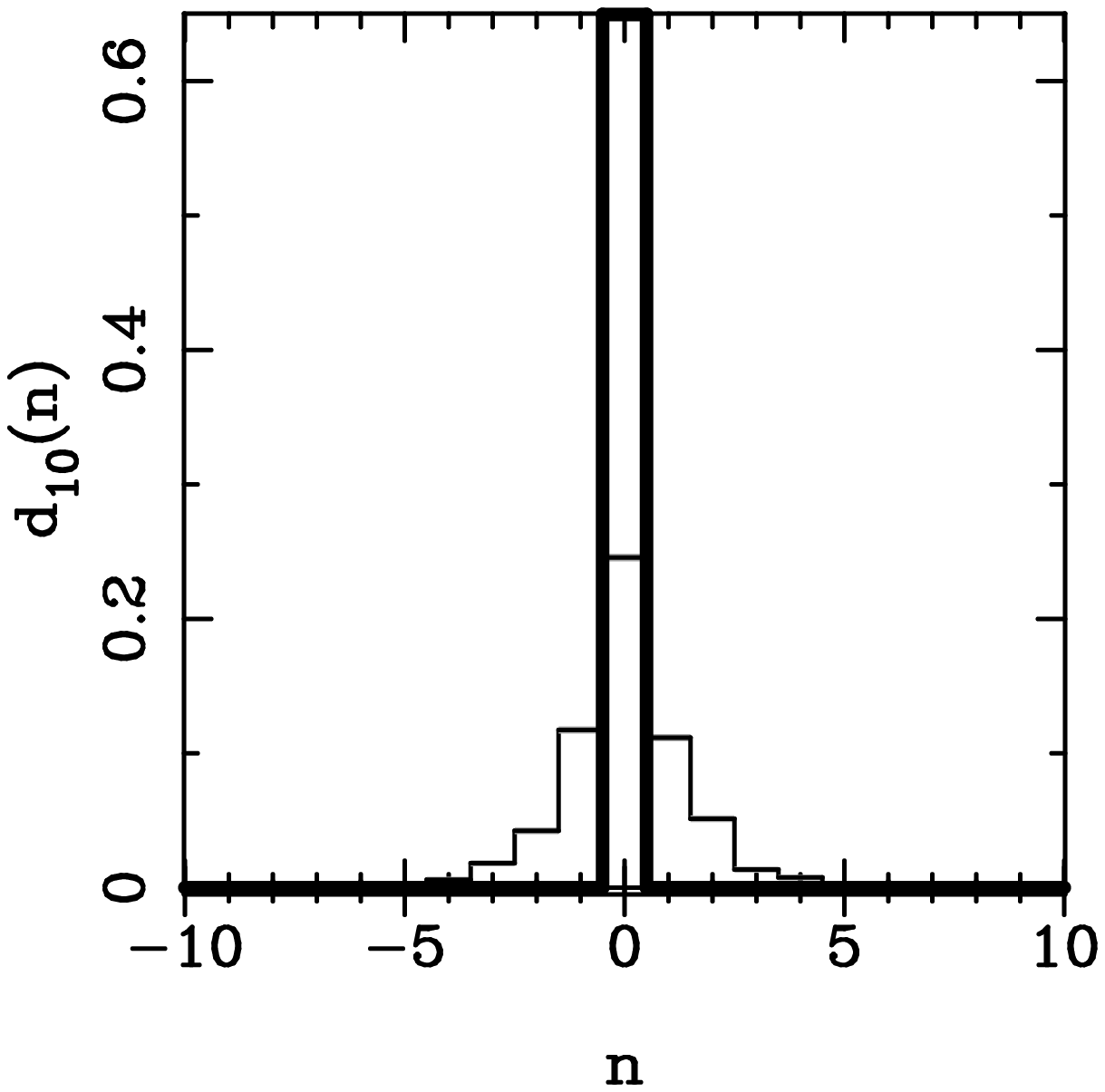}
\epsfxsize=.45\hsize\leavevmode\epsffile{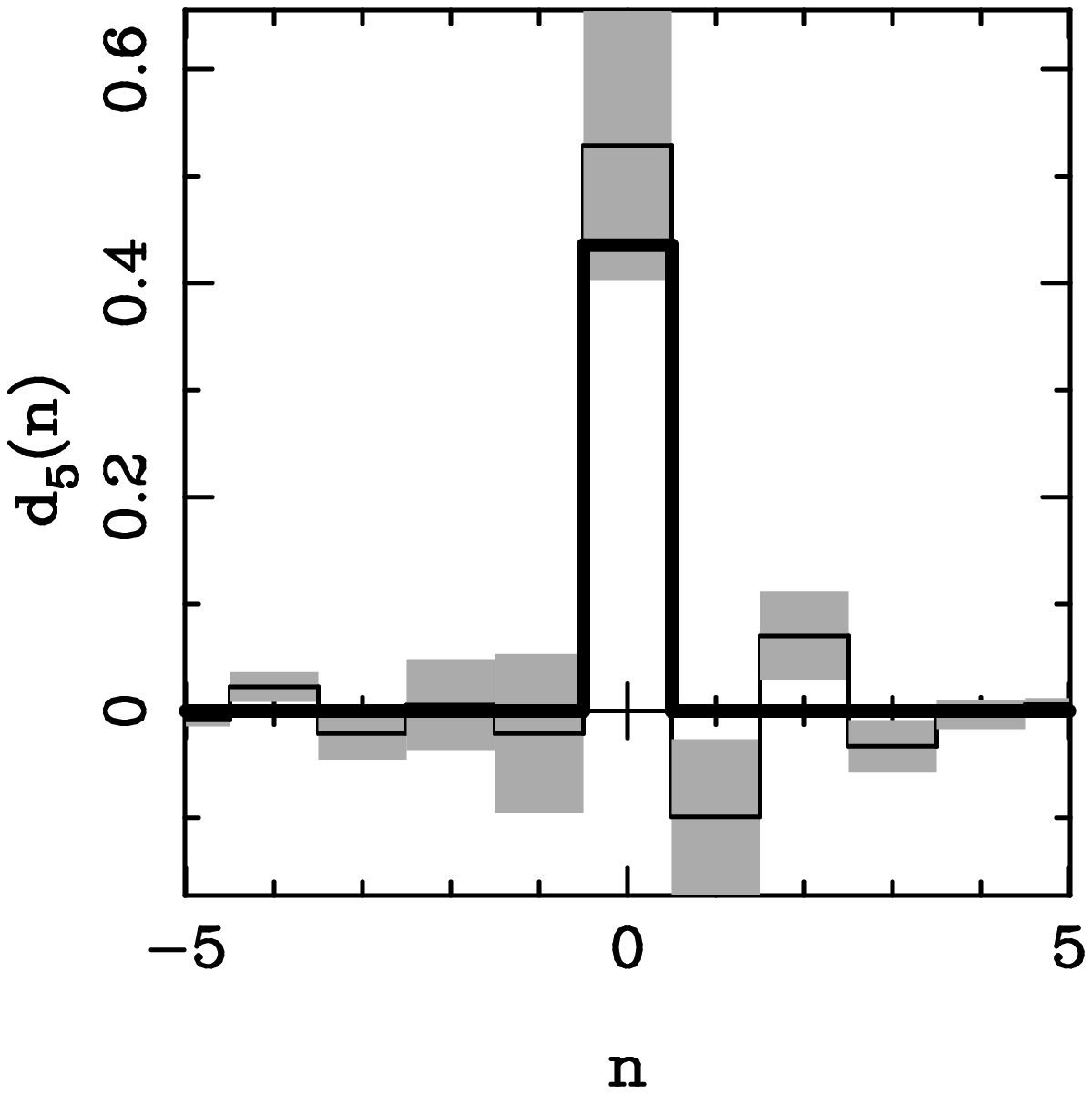}\end{center}
\caption{Similar to Fig.~\ref{f:G1}, but for $\eta=0.8$.
Results for the Gaussian-noise-dressed state reconstruction are shown 
on the left and for the bare 
state reconstruction on the right. In both simulations $5\times 10^6$
data samples were used. The non-unit quantum efficiency in the 
dressed-state case partially
smears out the correlation, which is recovered in the bare-state
reconstruction (right), however at the expense of increasingly-large
statistical errors.}\label{f:H1}\end{figure}

\begin{figure}[hbt]\begin{center}
\epsfxsize=.45\hsize\leavevmode\epsffile{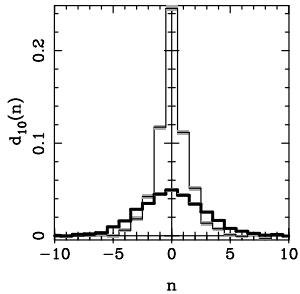}
\end{center}
\caption{Comparison of the correlation function, \protect{Eq.~(\ref{diff})},
for the Gaussian-noise-dressed
state of \protect{Fig.~\ref{f:H1}(left)}, shown by thin solid line with 
error bars in
gray shade, with that for two modes in uncorrelated
coherent states (thick solid line), having the same mean photon number 
$\bar n=10$ per mode.}
\label{f:HC1}\end{figure}

 Regarding measurement of the photon-number correlation $d_N(n)$ 
[Eqs.~(\ref{diff}) and~(\ref{thd})], 
comments similar to those made for the total photon-number 
probability $s(n)$ hold. Figure~\ref{f:G1}
presents the results of a simulation of the
correlation function for the twin-beam state with $N=\bar n=10$ and
unit quantum efficiency, whereas Fig.~\ref{f:H1} shows the results
of simulations with quantum efficiency $\eta=0.8$, once again, reconstructing
the correlation for both the
dressed-state and the bare-state cases. Here also, the non-unit quantum
efficiency in the case of dressed-state reconstruction partially smears out the
correlation, which is well recovered in the case of bare-state reconstruction.
In Fig.~\ref{f:HC1}, we compare the reconstructed correlation function for
the dressed state in Fig.~\ref{f:H1}(left) to that for two modes 
in uncorrelated coherent states, each having the same mean photon 
number $\bar n=10$ as the modes of the twin-beam state. 
One can see that, in spite of the 
detrimental effect of the non-unit quantum efficiency, the correlation for
the reconstructed dressed state is still stronger than that for the 
uncorrelated coherent states, the latter representing the standard 
quantum limit.
\section{Discussion}\label{s:discussion}
We have proposed a method for performing two-mode
optical-homodyne tomography of the twin beams produced from a 
nondegenerate optical 
parametric amplifier. The local oscillators
(LO's) needed for the homodyne
tomography are generated in the same parametric process as the twin beams,
and, therefore, are automatically matched to the signal and idler 
twin-beam modes.
In our self-homodyning method, the polarized central spectral
component at $\omega_{0}$ serves as the LO for a mode that consists 
of two sidebands at 
$\omega_{0}\pm\Omega$, and
the relative optical phase between the central component and the sidebands can be
varied. We have presented a theoretical description for both one-
and two-mode tomography, with main focus on measurement of the photon-number
distributions. For the signal mode alone, a thermal
distribution of photons is found, in agreement with the results of a recent
experiment
\cite{vasilyev97}. In the case of two modes, we have presented some selected 
Monte-Carlo simulations of the tomographic measurement of the joint
photon-number distributions, choosing realistic values for the quantum
efficiency of the photodetectors. In particular, we have analyzed the
feasibility of detecting photon-number oscillations and delta-like photon
correlation between the twin-beam modes. We have shown that for ideal 
photodetectors
such features can be clearly observed even with a small number of data
samples ($10^{6}$). However,
for realistic quantum efficiencies the oscillations are exhibited with
less contrast in the dressed-state reconstruction for the same number of
data samples. On the other hand, for
a tomographic measurement of the true output state of the OPA, more data 
samples are needed in order
to reduce the statistical errors. Our Monte-Carlo simulations show that
for a quantum efficiency of $\eta=0.9$, the
oscillations in the total photon number can be observed, 
even in the dressed-state reconstruction,
with as little as $10^6$ data samples, which makes such an experiment feasible.

We have also shown how the self-homodyning method can be used in detection of
the $\pm 45^{\circ}$-polarized modes, instead of the signal and idler modes. 
Since in a polarization-nondegenerate optical parametric amplifier these
modes are amplified independently, their joint photon-number 
distribution is factorized into a product of 
marginal distributions, each exhibiting even-odd oscillations in its 
photon number.

\par While the focus of our paper has been on the twin-beam state,
the self-homodyning approach can be applied in other instances as well. 
There are a number of mode-matching critical situations where it is
possible to mix the signal with another mode that underwent a similar
generation process. A key requirement in such situations would be 
the scanning of
the relative phase between the two modes. Among potential
applications are detection of the superposition (Schr\"odinger's cat)
states, and squeezed
states generated in optical fibers.
\section*{Appendix}
In this appendix we derive the joint probability distribution, 
Eq.~(\ref{pxy}),
of the output photocurrents for two-mode homodyne detection.
\par In the Fock representation, the state at the output of the NOPA
is given by Eq. (\ref{Psi}), namely,
\begin{eqnarray}
|\Psi\rangle=(1-\tau^2)^{1/2}\sum_{n=0}^{\infty}\tau^n|n,n\rangle\;.
\end{eqnarray}
Expanding the Fock state $|n\rangle$ in terms of the quadrature representation
$|x\rangle_{\phi}$ for each mode, one has
\begin{eqnarray}
&&|\Psi\rangle=
\sqrt{{2(1-\tau^2)}\over{\pi}}\int_{-\infty}^{+\infty}
dx\int_{-\infty}^{+\infty}dx'\,
e^{-x^2-{x'}^2}\nonumber\\ &&\times\sum_{n=0}^{\infty}{{\left[\tau
e^{-i(\phi+\psi)}\right]^n}\over{2^n n!}} H_n(\sqrt{2}x)
H_n(\sqrt{2}x') |x\rangle_{\phi}\otimes |x'\rangle_{\psi}\;,
\label{tw2}
\end{eqnarray}
where $H_n(x)$ denotes the Hermite polynomial of degree 
$n$. Using the following identity \cite{bateman}, which is 
valid for any complex number $z$, 
\begin{eqnarray}
&&\sum_{n=0}^{\infty}
{{\left({1\over2}z\right)^n}\over{n!}}H_n(x)H_n(x')\nonumber\\
&&= (1-z^2)^{-1/2}
\exp\left\{{{2xx'z-(x^2+x'^2)z^2}\over{1-z^2}}\right\}\;,
\end{eqnarray}
we can rewrite Eq. (\ref{tw2}) as
\begin{eqnarray}
|\Psi \rangle &=&\left[{{1-|\kappa|^2}\over{1-\kappa^2}}
\right]^{1/2} \sqrt{{2\over\pi}}
\int_{-\infty}^{+\infty}dx\int_{-\infty}^{+\infty}dx'\, 
|x\rangle_{\phi}\otimes |x'\rangle_{\psi}\nonumber\\
&&\times\exp
\left[{{4xx'\kappa-(x^2+{x'}^2)(1+\kappa^2)}\over{1-\kappa^2}}\right]
\;,\label{tw3}
\end{eqnarray}
where $\kappa=\tau\exp [-i(\phi+\psi)]$ (the choice of the branch 
for the square root in the normalization of the state vector (\ref{tw3}) 
gives only an overall phase factor that is irrelevant for
probabilities). Equation (\ref{tw3}) corresponds to the following joint
probability: 
\begin{eqnarray}
&&p(x,x';\phi,\psi)=\nonumber\\
&&\times {2\over{\pi |d_{\kappa}d_{-\kappa}|}}\exp\left[
-{{(x+x')^2}\over{d^2_{\kappa}}}-
{{(x-x')^2}\over{d^2_{-\kappa}}}\right]\;,\label{app:ideal}
\end{eqnarray}
where $d_{\kappa}^{2}\doteq|1+\kappa|^{2}/(1-|\kappa|^{2})$.
Non-unit quantum efficiency of the photodetectors is taken into account 
by evaluating the convolution of the ideal joint
probability in Eq.~(\ref{app:ideal}) with Gaussians for each mode of
variances given by
Eq.~(\ref{vareta}). This immediately leads to Eq.~(\ref{pxy}).
\acknowledgments
This work was supported in part by the U. S. Office of Naval Research.

\end{document}